\documentclass[a4paper,fleqn]{cas-sc}

\usepackage[numbers,sort&compress]{natbib}
\usepackage{soul}
\usepackage{subfigure}
\usepackage{tikz}
\usepackage{adjustbox}
\usepackage{lipsum}
\usepackage{setspace}   

\def\tsc#1{\csdef{#1}{\textsc{\lowercase{#1}}\xspace}}
\tsc{WGM}
\tsc{QE}
\tsc{EP}
\tsc{PMS}
\tsc{BEC}
\tsc{DE}

\begin{document}
\let\WriteBookmarks\relax
\def\floatpagepagefraction{1}
\def\textpagefraction{.001}
\shorttitle{Powder Technology}
\shortauthors{Islas et~al.}

\title [mode = title]{CFD Simulations of Turbulent Dust Dispersion in the 20L Vessel using OpenFOAM}                                    



\author[1]{Alain {Islas}}[]


\address[1]{Department of Energy, University of Oviedo - 33203 Gijón, Asturias, Spain}

\author[1]{Andrés {Rodríguez-Fernández}}[]

\author[2]{Covadonga Betegón}[%
   ]


\address[2]{Department of Construction and Manufacturing Engineering, University of Oviedo - 33203 Gijón, Asturias, Spain}

\author[3]{Emilio Martínez-Pañeda}[]

\address[3]{Department of Civil and Environmental Engineering, Imperial College London - London, SW7 2AZ, United Kingdom}

\author[1]{Adrián Pandal}[orcid=0000-0001-6006-2199]
\cormark[1]

\cortext[cor1]{Corresponding author:}


\begin{abstract}
Dust explosions are among the most hazardous accidents affecting industrial facilities processing particulate solids. Describing the severity parameters of dust clouds is critical to the safety management and risk assessment of dust explosions. These parameters are determined experimentally in a 20L spherical vessel, following the ASTM E1226 or UNE 14034 standards. Since their reproducibility depends on the levels of turbulence associated with the dust cloud, a computational model of the multi-phase (gas-solid) flow is used to simulate the dispersion process with the open-source CFD code OpenFOAM. The model is successfully validated against experimental measurements from the literature and numerical results of a commercial CFD code. In addition, this study considers the impact of particle size on the turbulence of the carrier phase, suggesting that particles attenuate its turbulence intensity. Moreover, the model predicts well the formation of a two-vortex flow pattern, which has a negative impact on the distribution of the particle-laden flows with $d_{\text{p}}\leq$ 100 $\mu\text{m}$, as most of the particles concentrate at the near-wall region. Contrarily, an improved homogeneity of dust cloud is observed for a case fed with larger particles ($d_{\text{p}}=$ 200 $\mu\text{m}$), as the increased inertia of these particles allows them to enter into the re-circulation regions. 





\end{abstract}


\maketitle

\doublespacing

\section{Introduction}
\label{Section:Introduction}

Powdered materials play an important role in modern chemical and process industries. It is estimated that three-quarters of all raw materials used in these industries and half its products are in particulate form \cite{ogle2016dust}. Although powdered materials are well suited for transporting, handling and storing operations, under right circumstances, these can behave as combustible dusts, leading to the development of dust explosions \cite{SCHULZ201937}. These occur when dust is airborne, oxygen is present and there is a source of ignition \cite{cheremisinoff2014dust}. According to Eckhoff \cite{eckhoff2003dust}, a dust cloud is easier to ignite and burns more violently the smaller the dust particles are. When explosive combustion of dust clouds takes place inside process equipment, the pressure inside may rise rapidly often producing catastrophic damage to facilities and causing large-scale loss of life \cite{csb2006investigation}. For example, a total of 281 dust explosions took place in the United States between 1980 and 2005, resulting in 119 fatalities and 718 injuries \cite{CLONEY201933}. More recently, in only the first half of 2020, 26 dust explosions were reported worldwide \cite{DSS_2020}. \\


To determine whether a dust is hazardous, standard test methods \cite{ASTME1226,EN14034,ISO6184-1} measure the potential of dust clouds to explode by determining its deflagration index, $K_{\text{st}}$. Following this approach, laboratory test data allow to predict conservatively the consequences if the same mixture accidentally explodes in an industrial plant through the cubic root law. In this relationship the deflagration index is directly connected to the maximum rate of pressure rise, ${(\textit{dP/dt})_{max}}$, and the considered volume, $V$, as presented in Eq. (\ref{Equation:cubic_root_law}) \cite{bartknecht1971brenngas,bartknecht1980explosionen,DIBENEDETTO2007303}. Since this individual value can predict the dust explosion violence, it is widely used as a key parameter of vent sizing to protect industrial equipment (e.g. silos, conveyors or rolling mills) \cite{NFPA68,TASCON2011717,TORNO2020310,THEIMER1973137}.\\

\begin{equation}
    K_{\text{st}} = \left(\frac{dP}{dt}\right)_{max}V^{1/3}
    \label{Equation:cubic_root_law}
\end{equation}

In these standardized tests, a determined amount of dust (i.e. the needed one to fulfil chamber concentration) is stored in a container. This sample is injected into vessel volume by means of a pressure-driven flow and is dispersed as it flows through a nozzle device. After an ignition delay time,  $t_{\text{d}}$, a turbulent dust cloud is created inside the test chamber. This cloud is immediately burned (typically using pyrotechnical ignitors) and the explosion pressure history is registered. This procedure is repeated for a range of concentrations, from which both $K_{\text{st}}$ and $P_{\text{max}}$, the maximum registered over-pressure, are derived.\\

Historically, the $1\text{m}^3$ pressure-resistant vessel was conceived at the Bergbau-Versuchsstrecke (BVS) in Germany under the leadership of Dr. Wolfgang Bartknecht \cite{BARTKNECHT1966}. Due to its fair reproducibility of dust explosions, it was established as standard instrument for dust explosion testing for many years since its first ISO regulation in 1985 \cite{ISO6184-1}. However, due to its design size, an extensive amount of dust is required and using it in a normal laboratory becomes complicated. Therefore, since 1988 the 20L spherical vessel proposed by Siwek \cite{siwek197720} became the preferred experimental apparatus for determining explosive parameters, overcoming the major drawbacks of the $1 \text{m}^3$ vessel and requiring a significantly lower testing time. Nonetheless, the $1\text{m}^3$ vessel remains as the "gold standard" in dust explosion testing, allowing to verify any result that seems suspicious in the 20-L one.\\

A critical aspect for the truthfulness of the 20L vessel is to achieve the same levels of turbulence for the air-dust mixture at the end of the dispersion process that occurred in the $1\text{m}^3$ vessel. As shown by Bartknecht \cite{bartknecht1989dust}, turbulence exerts a significant influence on the explosion characteristics, as larger $K_{\text{st}}$ values are obtained with smaller $t_{d}$. Experiments at that time suggested that the same turbulence conditions existed when $t_{\text{d}}=60 \,\text{ms}$ and $t_{\text{d}}=600 \,\text{ms}$, for the 20L sphere and the $1 \text{m}^3$ chambers, respectively. Later, Pu et al. \cite{pu1991turbulence} and Dahoe et al. \cite{dahoe2001transient} measured turbulent flow properties in the 20L chamber during the ignition delay time. They found that different turbulence levels existed between the two vessels, hence questioning the validity of the cubic-root law. Van der Wel et al. \cite{van1992interpretation} performed similar measurements in the $1\text{m}^3$ chamber. All of these studies suggested that setting a modified ignition delay time of about 165-200 ms for the $1\text{m}^3$ one would have a much better agreement in terms of turbulence levels. In 2007, Proust et al. \cite{PROUST2007599} tested a wide range of twenty-one different powders in both apparatuses. They found that $P_{\text{max}}$ was systematically lower in the 20L vessel while for $K_\text{st}$ only six powders shown a standard deviation between the two chambers lower than the 5\%, exceeding the 20\% in most cases, which reinforced conclusions from previous studies.\\

Another key assumption made in these tests is that homogeneous dust clouds are created inside the test chambers. Specific nozzle devices were designed to create fairly uniform dust clouds, being the so called "rebound" and "annular" nozzles the standard dispersion devices. While the perforated annular nozzle was able to generate a quite uniform dust cloud, a significant fraction of solid particles is trapped inside the nozzle \cite{DISARLI2015204}. Besides, measurements with the rebound nozzle have evidenced that the dust cloud is particularly concentrated in the wall region \cite{vizcaya2018cfd,DIBENEDETTO2013cfd}, creating zones of low dust concentration near the ignitors \cite{di2014cfd} and therefore affecting the flame propagation during the combustion of the dust particles. Despite their difficulty to be performed,  a few authors have developed experimental techniques and vessel modifications to perform internal measurements which have confirmed these trends \cite{KALEJAIYE201046,du2015visualization}. Sanchirico et al. \cite{SANCHIRICO2015203} studied the effect of varying the injection pressure and compared particle size distributions before and after dispersion proving that particle breakage occurred at any conditions for every investigated dust. Likewise, Computational Fluid Dynamics (CFD) studies have also been applied to achieve an accurate description of particle behavior. Murillo et al. \cite{murillo2016cfd} simulated the dispersion of wheat starch powder and suggests that particle fragmentation is more dependent on the flow dynamics compared to other secondary mechanisms, like the collisions against the nozzle walls. Similarly, Vizcaya et al. \cite{vizcaya2018cfd} tested the same dust and found that according to the post-dispersion granulometric analysis the particle mean diameter experienced a reduction of 69\%, revealing an important degree of particle fragmentation during dispersion. When comparing the influence of different nozzles, particle fragmentation is more prone to occur when employing the rebound nozzle, as this is promoted by higher turbulence levels than to the acting mechanism of the outlet valve previously pointed out \cite{KALEJAIYE201046}.\\

Increasing computational capabilities suggest that CFD models could be an effective tool to simulate the hazardous potential of dust clouds behaviour not only in the standardized experimental vessels but also in actual industrial facilities. Within the framework of an EU project, researchers developed DESC (Dust Explosion Simulation Code) \cite{SKJOLD2007291}, which later became a sub-module (FLACS-DustEx) in Gexcon's commercial CFD software for simulating gas and dust explosions \cite{FLACS}. However, the evaluation of $K_{\text{st}}$ still relies on the same cubic-root law, which is valid for specific turbulence levels and that might not be equal for every industrial application. Then, a more robust model is advisable.\\

Although extensive research has been conducted along the recent history, dust explosions in the process industries are still a major issue and need better methods for predicting real dust cloud generation, ignition, combustion, and flame propagation processes. This work presents the initial step to construct a reliable engineering tool for the simulation of large-scale dust explosions in specific industrial geometries. To this end, it is fundamental the prediction of turbulence levels, degrees of dust dispersion, and distributions of dust concentrations encountered inside the vessels to provide an accurate basis for the subsequent dust explosion modeling. Thus, in the present paper, the dust dispersion process is considered and studied by means of the standardized 20L sphere (equipped with the rebound nozzle and proceeding according to the ASTM E1226 standard \cite{ASTME1226}). In contrast with previous CFD studies on the 20-L vessel turbulent dispersion, conducted using commercial codes (ANSYS Fluent® \cite{DIBENEDETTO2013cfd,LI2020118028,RAY2020321,WANG2019509} or STAR-CCM+® \cite{MURILLO2013103,murillo2016cfd}), this work relies on the use of Open-FOAM (Open-source Field Operation And Manipulation) \cite{WELLER1998}, a CFD open-source C++ library that has gained wide recognition in academic, research and industrial sectors \cite{LYSENKO2013408,TAVAKKOL2021110582}. Results obtained with OpenFOAM are compared with both experimental and CFD predictions available in the literature. The shown great performance of the model encourages its further development towards the accomplishment of the final goal.\\

\section{Model description}

From the wide availability of solvers in OpenFOAM 8, \texttt{coalChemistryFoam} is chosen, which is a multi-phase transient solver for compressible flow. The multi-phase flow interaction (gas-particles) is modeled by an Eulerian-Lagrangian approach. It also includes particle advanced models as well as advanced chemistry and combustion models. Rather, no reactive features of the solver have been used in this study, as research is focused on the dispersion stage only. The fluid flow is described by the compressible form of the unsteady Reynolds-Averaged Navier-Stokes (URANS) equations, Eqs. (\ref{Equation:mass_transport},\ref{Equation:momentum_transport}) and a standard $k$-$\varepsilon$ turbulence model \citep{launder1983numerical}, Eqs. (\ref{Equation:k_transport},\ref{Equation:epsilon_transport}), has been used, including the effect of a source term due to interaction between phases. According to Elghobashi's work \cite{elghobashi1994predicting}, volume fractions considered in this study fit a two-way coupling regime, hence making possible to neglect the effect of particle-particle interaction.

\begin{equation}
    \frac{\partial \bar{\rho}}{\partial t}+\frac{\partial}{\partial x_i}\left(\bar{\rho} \widetilde{u_i}\right)=0
    \label{Equation:mass_transport}
\end{equation}
\begin{equation}
    \frac{\partial}{\partial t}\left(\bar{\rho} \widetilde{u_i}\right) + \frac{\partial}{\partial x_j}\left(\bar{\rho} \widetilde{u_i} \widetilde{u_j}\right) = - \frac{\partial \bar{p}}{\partial x_j} + \frac{\partial \bar{\tau}^{ij}}{\partial x_j}+\frac{\partial}{\partial x_j}\left(-\bar{\rho} \widetilde{u_i^{'} u_j^{'}}\right) +\bar{\rho} g_i + S_{i}
    \label{Equation:momentum_transport}
\end{equation}
\begin{equation}
       \frac{\partial}{\partial t}\left(\bar{\rho} k\right)+\frac{\partial}{\partial x_i}\left(\bar{\rho} \widetilde{u_i}k\right) = \frac{\partial}{\partial x_j}\left[\left(\mu + \frac{\mu_t}{\sigma_k}\right)\frac{\partial k}{\partial x_j}\right] + P_{k} +P_{b}-\bar{\rho} \varepsilon - Y_{M} + S_{k}
       \label{Equation:k_transport}
\end{equation}
\begin{equation}
    \frac{\partial}{\partial t}\left(\bar{\rho} \varepsilon\right)+\frac{\partial}{\partial x_i}\left(\bar{\rho} \widetilde{u_i}\varepsilon\right)=\frac{\partial}{\partial x_j}\left[\left(\mu+\frac{\mu_t}{\sigma_{\varepsilon}}\right)\frac{\partial \varepsilon}{\partial x_j}\right] + C_{1\varepsilon}\frac{\varepsilon}{k}\left(P_{k}+C_{3\varepsilon}P_{b}\right)-C_{2\varepsilon}\bar{\rho}\frac{\varepsilon^2}{k}+S_{\varepsilon}
    \label{Equation:epsilon_transport}
\end{equation}
\begin{equation}
    \frac{\partial}{\partial t}\left(\bar{\rho} \widetilde{h_0}\right) + \frac{\partial}{\partial x_i}\left(\bar{\rho} \widetilde{u_i} \widetilde{h_0}\right) = \frac{D \bar{p}}{D t} - \frac{\partial \bar{q_i}}{\partial x_i} + \overline{\tau^{ij}\frac{\partial u_i}{\partial x_j}}
    \label{Equation:energy_transport}
\end{equation}

Lagrangian phase is solved by applying Newtons 2\textsuperscript{nd} law to the particles, rendering the following force balance:

\begin{equation}
    \frac{du_{p_i}}{dt}=\frac{18\mu}{\rho d_{p}^{2}}\frac{C_{D}\text{Re}_{p}}{24}\left(u_i-u_{p_{i}}\right)+g_i\left(1-\frac{\rho}{\rho_p}\right)+F_{\text{other}}
    \label{Equation:particle_force_balance}
\end{equation}

     \[ C_D = \begin{cases} 
      0.424 & \text{Re}_{p}> 1000 \\
      \frac{24}{\text{Re}_{p}}\left(1+\frac{1}{6}\text{Re}_{p}^{2/3}\right) & \text{Re}_{p}\leq 1000
      \label{Equation:Drag_coefficient}
   \end{cases}
\]

\noindent where the drag coefficient $C_D$, is taken from Putnam \cite{putnam1961integratable}, and the particle Reynolds number is defined as \linebreak $\text{Re}_{p}=\rho d_{p}|u_i-u_{p_i}|/\mu$. This correlation is valid for spherical particles and is suitable for high Reynolds numbers \cite{crowe2011multiphase}. In this study, the term $F_{\text{other}}$ is set to zero, as the dominant forces for micron particles with a low fluid-to-particle density ratio are mainly drag, gravity and buoyancy.\\

Each one of these computational parcels represents a physical entity by a cluster of particles that are assumed to share the same properties. Hence, all the conservation equations are scaled by the number of particles present in the parcel. This technique is called Discrete Parcel Method (DPM) \cite{haervig2017adhesive} and is broadly used in CFD to reduce the computational burden. The amount of particles grouped in each parcel must be set not only to ensure that parcels represent particles properties fairly well but also that the overall motion of the discrete phase is statistically significant \cite{PICO2020638}.\\


The turbulence effect on particle trajectories is accounted by means of a stochastic dispersion approach, in which a fluctuating velocity component is randomly sampled from a Gaussian distribution with $\sigma=\sqrt{2k/3}$ (assuming isotropic turbulence) and that is added to $u_{i}$. Another important aspect of this approach is the manner to calculate the time over which a particle interacts with a turbulent eddy. Following an approach similar to Gosman and Ioannides \cite{gosman1983aspects}, here the "interaction" time, $t_{int}$, is computed as:

\begin{equation}
    t_{int} = \min\left(k/\varepsilon,l_{e}/|u_{i}-u_{p_{i}}| \right)
    \label{Equation:Particle_interaction_time}
\end{equation}

\noindent where $l_{e}$ is the dissipation length scale. Finally, due to the compressibility effect of the flow, the particles undergo a temperature change during the dispersion process. This can be found by applying an energy balance to the lagrangian phase:

\begin{equation}
    m_{p_i}C_p\frac{dT_{p_i}}{dt}=\pi d_p^2 h \left(T_{\infty}-T_{p_i}\right)
    \label{Equation:Paricle_energy_balance}
\end{equation}

\noindent where the heat transfer coefficient, $h$, is taken from Ranz \& Marshall \cite{ranz1952evaporation}. Internal temperature gradients are not considered.


\section{Numerical discretization}

\subsection{Geometry and computational grid}

\begin{figure}
	\centering
		\includegraphics[scale=.40]{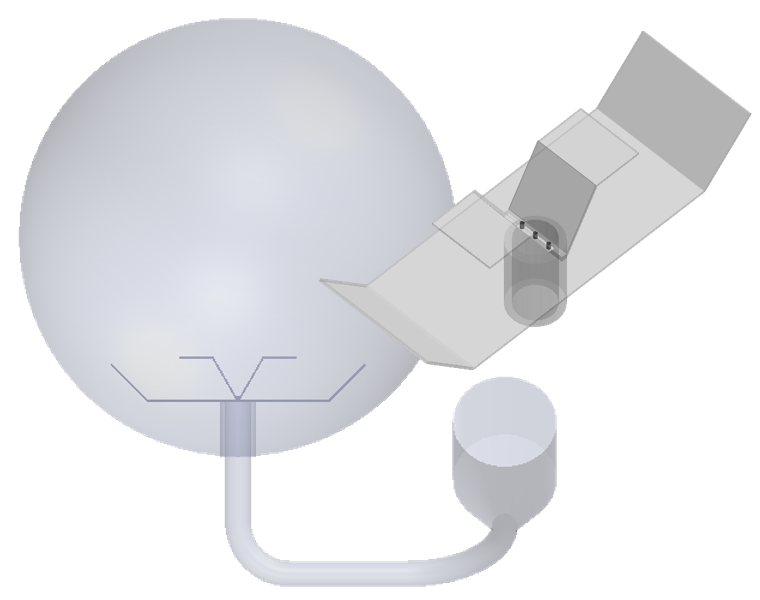}
	\caption{Computational domain of the 20L spherical vessel for dust explosions and detail of the rebound nozzle.}
	\label{Figure:Geometry}
\end{figure}

The computational domain has been modeled to represent the equipment as closely as possible to the standards specifications available in the literature, as they are the ones also followed by manufacturers. The canister is considered to have a bottle-shaped geometry with a volume of 0.6 L, while it is connected to the sphere vessel by means of a pipe with two elbows, see Fig. \ref{Figure:Geometry}. This is placed in a skewed plane relative to the mid-plane of the sphere. The model is equipped with the rebound nozzle, whose dimensions were taken from the ASTM E1226 standard. For the sake of simplicity, inside the sphere pyrotechnic ignitors were omitted since combustion is not analyzed. However, it was decided that modeling the actual dimensions and location of the canister and connecting pipe were an important aspect to reproduce precisely the actual injection process of particles within the sphere. \\

\begin{figure}
    \centering
    \includegraphics[width=0.43\textwidth]{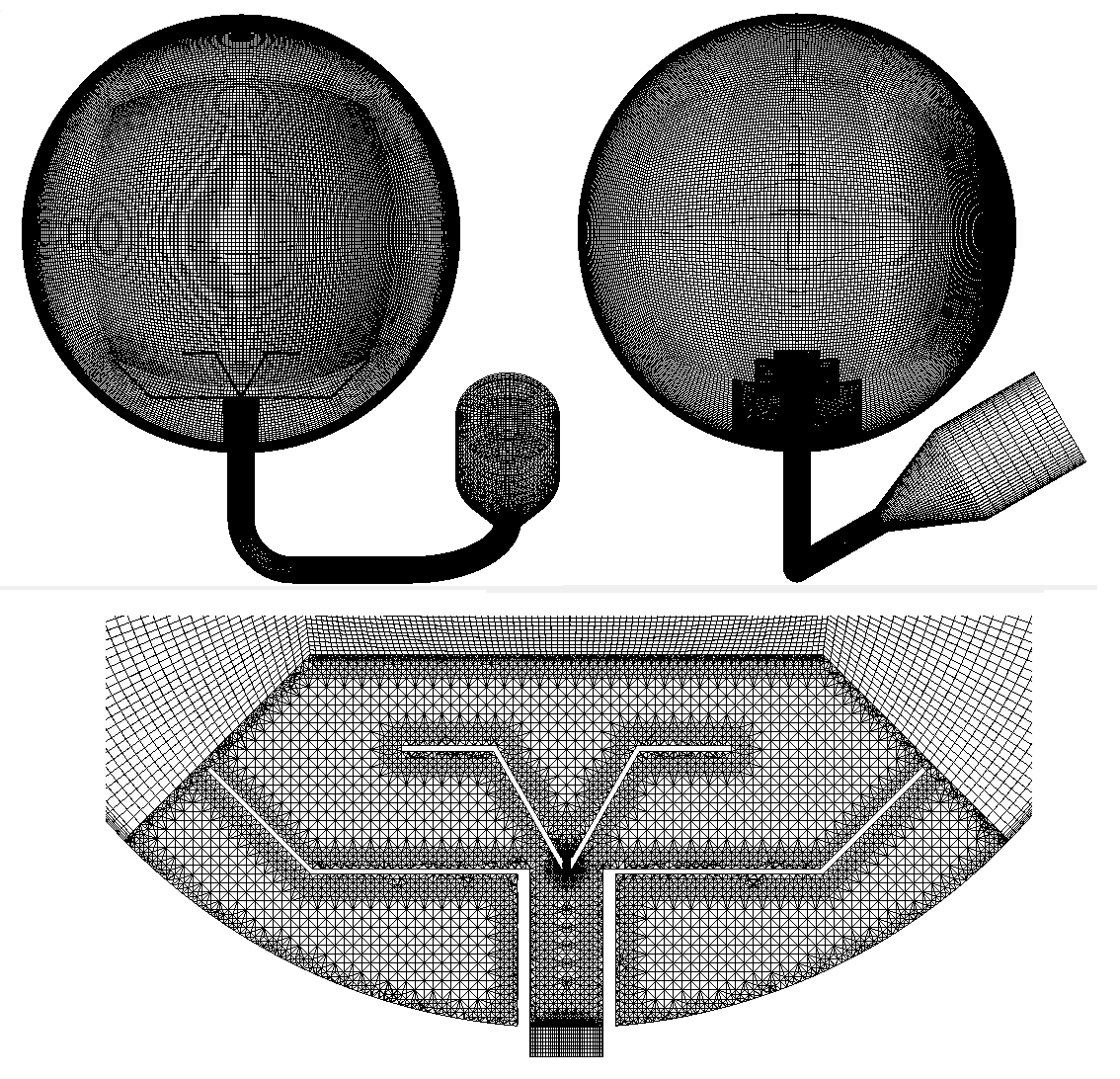}
    \caption{Computational grid with hybrid layout including detail of tetrahedrons region surrounding the nozzle.}
    \label{Figure:mesh}
\end{figure}

The computational grid was generated in ANSYS ICEM\textsuperscript{\textregistered} and it is comprised of hexahedron elements for the canister and pipe regions, see Fig. \ref{Figure:mesh}. The 20L vessel was discretized with a hybrid combination of hexahedron and tetrahedron elements; using a 3D C-grid for the top and sides of the sphere and tetrahedrons for the region surrounding the nozzle. Transition between elements was achieved with pyramid elements.\\

This newly-purposed method provides a high orthogonality in 80\% of the fluid region of interest, while keeping the advantages of automatic grid generation in regions that are not suitable for manual meshing. Moreover, to validate that the external grid is in conformance with the requirements of the solver, the mesh metrics lie within the suitable ranges of OpenFOAM's quality indices. The latter are calculated by the built-in \texttt{checkMesh} utility, and validated that the mesh geometry and topology were not corrupted. These are listed in Table \ref{TAB:1}.\\



\subsection{Case description and boundary conditions}

 As per the standard test procedure, the flow inside the domain is considered as quiescent air, hence velocity and turbulence fields are set accordingly. The air is assumed as an ideal gas and the dust is initially placed in the frustum section of the canister, mimicking a real deposition of the dust sample into the container.
 The fluid region is patched according to the prescribed absolute pressures of the standard, this is 21 bar and 0.4 bar for the canister and sphere regions, respectively.\\
 
 Moreover, in the real experiment the equipment is covered by a double water-jacket that provides a cooling effect. This is particularly useful to avoid a temperature increase between a series of dust explosion tests. Hence, the wall temperature is set to 293 K, whereas the nozzle device is considered as adiabatic. Finally, a no-slip shear condition is defined for all the wall region while the particle-wall interaction is specified with elastic reflections for both normal and tangential directions. A summary is listed in Table \ref{TAB:2}.

\subsection{Numerical schemes and grid dependency test}

Eqs. (\ref{Equation:mass_transport}-\ref{Equation:energy_transport}) are discretized utilizing a first-order upwind scheme for convective terms and a second-order central differencing scheme for diffusion terms. Gradient terms are evaluated using a cell-limited scheme with cubic interpolation. Temporal discretization was calculated using a first-order Euler scheme with an adaptive time-stepping method to satisfy a Courant number of 5. The velocity-pressure coupling is handled by the PIMPLE algorithm with 3 correctors for each time step. Residuals were set to $10^{-8}$ and $10^{-12}$ for continuity-pressure and momentum-turbulence equations, respectively.\\

For the Lagrangian phase, particle velocities are found by solving Eqs. (\ref{Equation:particle_force_balance},\ref{Equation:Paricle_energy_balance}) using an Euler integration method with a limiting Courant number of 0.3. This guarantees the stability of the coupled solution between the Eulerian and Lagrangian phases, at the same time ensuring that computational parcels do not travel across more than one cell per time step.\\


Previous CFD studies on the dispersion of dust particles in the 20L sphere have employed tetrahedral and polyhedral meshes \cite{di2014cfd,MURILLO2013103}; however none of them have considered those containing hexahedral elements, which are often referred as the most suitable meshing elements that provide high accuracy, while keeping high ortogonality and a reduced number of cell faces \cite{baker2005mesh}. As the performance of a hybrid mesh of the 20L spherical vessel (composed nearly of 80\% of hexahedral elements) has not been previously studied, an analysis devoted to grid independence is considered here.\\

\begin{figure}
    \centering
    \includegraphics[width=\textwidth]{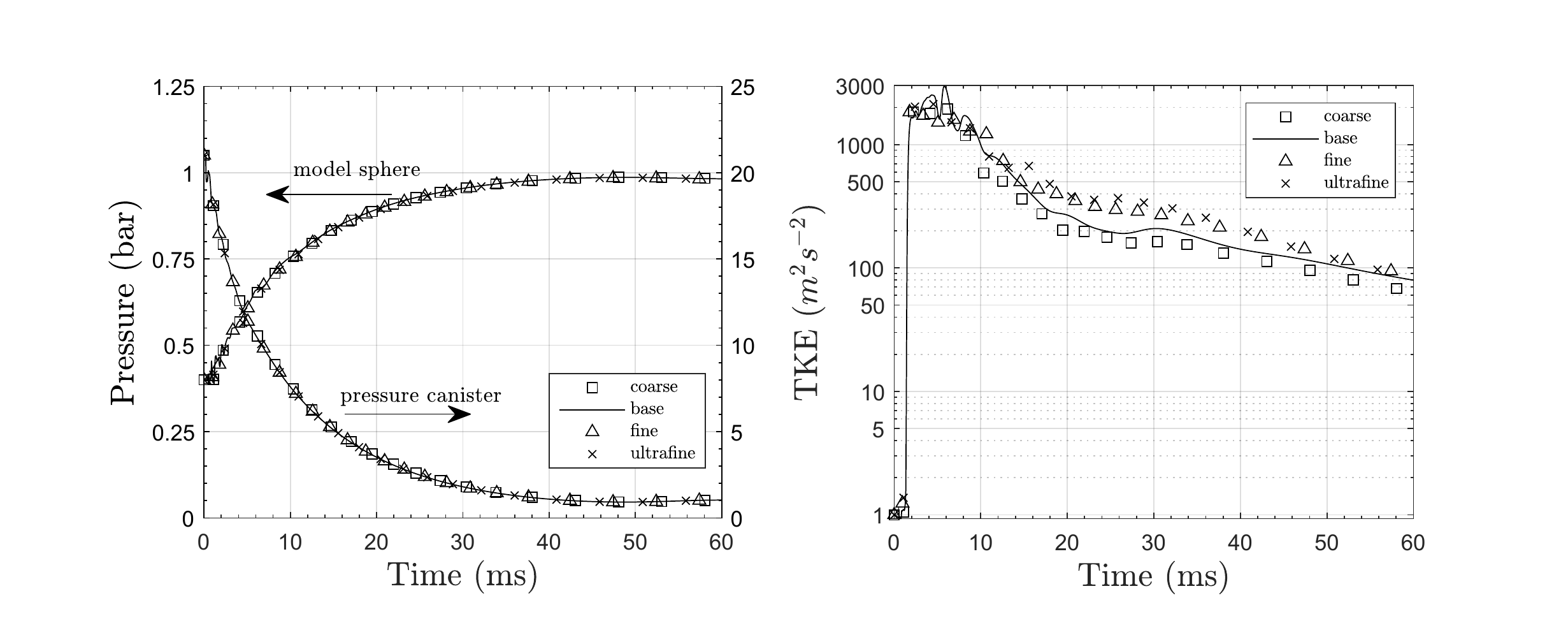}    
    \centering \caption{Comparison of the temporal trend of the pressure in the sphere and canister (left) and, evolution of the TKE in the sphere (right) for the different grid sizes evaluated.}
    \label{Figure:GCI_variables}
    
\end{figure}

Grid independence was checked with four grids, namely: ultra-fine (7.5M), fine (3.25M), base (1.62M), and coarse (0.83M). The cell count was calculated to approximate a grid refinement ratio, $r$, in the vicinity of 1.3. The performance of the grids was calculated using the variables of interest to the flow, the pressure and turbulent kinetic energy (TKE) records. Fig. \ref{Figure:GCI_variables} shows the temporal trends of the pressure and TKE of the dust-free flow simulation. As appreciated, the pressure profiles show an excellent agreement for all the four grids, suggesting that the pressure has converged since the employment of the coarsest grid. Conversely, the TKE profiles exhibit a clear cut sensitivity to the grid size. The profiles of the fine and ultra-fine meshes overestimate the decay of turbulence with respect to that of the base grid. Contrarily, the latter is underestimated when the coarse grid is employed.\\

Moreover, the numerical error introduced by the spatial discretization was calculated following the GCI method, described by Celik et al. \cite{celik2008procedure}. Here, the characteristic grid size, $h$ (in mm), for the four grids  is calculated as the cube-root of the average cell volume, particularly: $h_{1}=1.4, h_{2}=1.85, h_{3}= 2.33$ and $h_{4}=2.92$, where the subscripts 1, 2, 3 and 4 correspond to ultra-fine, fine, base and coarse, respectively. Two sets were considered for the calculation of the GCI, first the one consisting of grids 4, 3, 2 and second, another set consisting of grids 1, 2, 3. The values considered in the calculation of the GCI are listed in Table \ref{Table:GCI_results}. On the one hand, the numerical error in the calculation of the pressure field is very small, as the grid convergence index is $0.11\% $ and $0.007\% $ for the first and second grid sets, respectively. These values confirm that the pressure field has already converged. Furthermore, in the second grid set the extrapolated value $\phi^{21}_{\text{ext}}$ already matches the pressure value obtained with the ultra-fine grid $\phi_{1}$ suggesting that further refinement will have an insignificant effect on this field. On the other hand, the numerical error in the calculation of the TKE is slightly higher, as the grid convergence indexes are $6.93\%$ and $5.33\%$ for the first and second grid sets, respectively. The sensitivity of the TKE to the grid size is related to the turbulent eddy viscosity, which is strongly affected by flow history effects and that is difficult to match identically among simulation runs. However, these errors rely on a local TKE value and can be considered acceptable as far as the most important features of the flow are captured, such as the periods of turbulence build up and decay. Moreover, in all cases the solutions of both pressure and TKE are in the asymptotic range of convergence.\\

\begin{figure}
    \centering
    \includegraphics[width=0.45\textwidth]{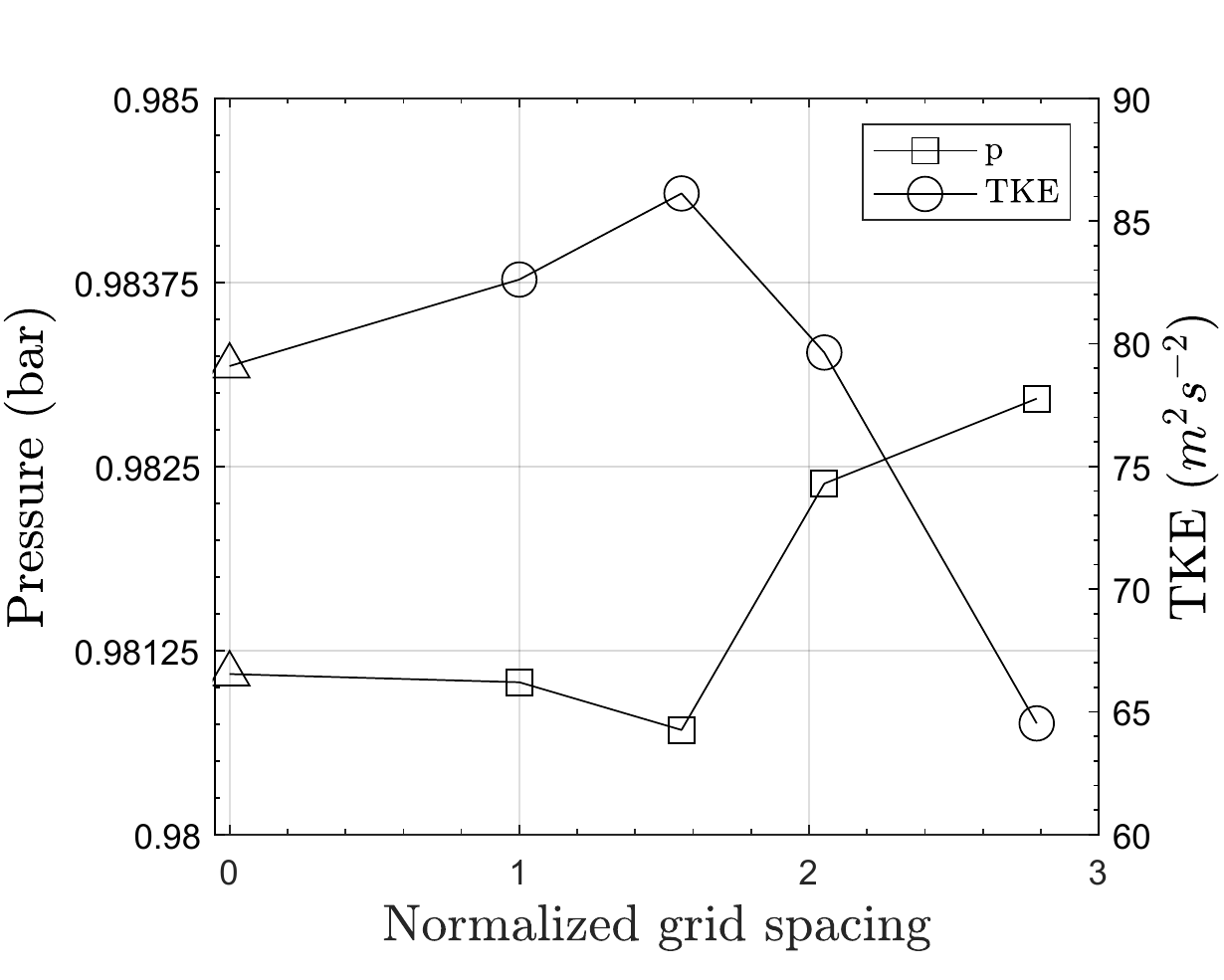}
    \caption{Rate of convergence of the pressure and TKE in the sphere at a time $t=60$ ms with respect to the grid spacing. Triangles represent the corresponding extrapolated values.}
    \label{Figure:Convergence_grid_spacing}
\end{figure}

Another remarkable aspect is that for the first grid set, a monotonic convergence for both pressure and TKE is observed, while the second grid set exhibits oscillatory convergence, see Fig. \ref{Figure:Convergence_grid_spacing}. This behavior suggests that further refinements up to that comparable to an infinitesimally small grid spacing (at Richardson's extrapolation) will approach to an intermediate value of the already computed TKE values. Hence, for the rest of the simulations, it is concluded that the grid that provides acceptable accuracy at a moderate computational cost is the base grid consisting of 1.62M cells, which also provides the TKE value that is more approximate to the extrapolated value of the second grid set.\\



\section{Results and discussion}
\subsection{Validation of the CFD model}

To validate the model, CFD results obtained with OpenFOAM are initially compared to the experimental measurements of Dahoe et al. \cite{dahoe2001transient}. In their study, the RMS velocity fluctuations of the dust-free flow were monitored using laser Doppler anemometer while the pressure change inside the sphere chamber and canister was recorded with piezo-electric transducers. First, a theoretical prediction of the final pressure that is reached in the sphere can be estimated if the initial and final temperatures in both canister and 20L vessel are equal. Following the ideal-gas law:
 
 \begin{equation}
     p_{f}=\frac{p_{c_i}V_{c}+p_{s_i}V_{s}}{V_{c}+V_{s}}
     \label{Equation:pressure_equilibrium}
 \end{equation}
 
 \noindent the pressure at the end of the dispersion process reaches exactly 1 bar in the domain, where $V_{c}$ and $V_{s}$ are the volumes of the canister and sphere vessel, respectively.\\

\begin{figure}
    \centering
    
    \subfigure[(Dust-free flow) Comparison of the temporal trends of the pressure inside the sphere and canister (left) and evolution of the RMS velocity fluctuations up to 1s in the sphere (right) between experimental measurements (Dahoe et al. \cite{dahoe2001transient}) and CFD results obtained with OpenFOAM in this study. \label{Subfigure:Validation_Dahoe}]{\includegraphics[width=1\textwidth]{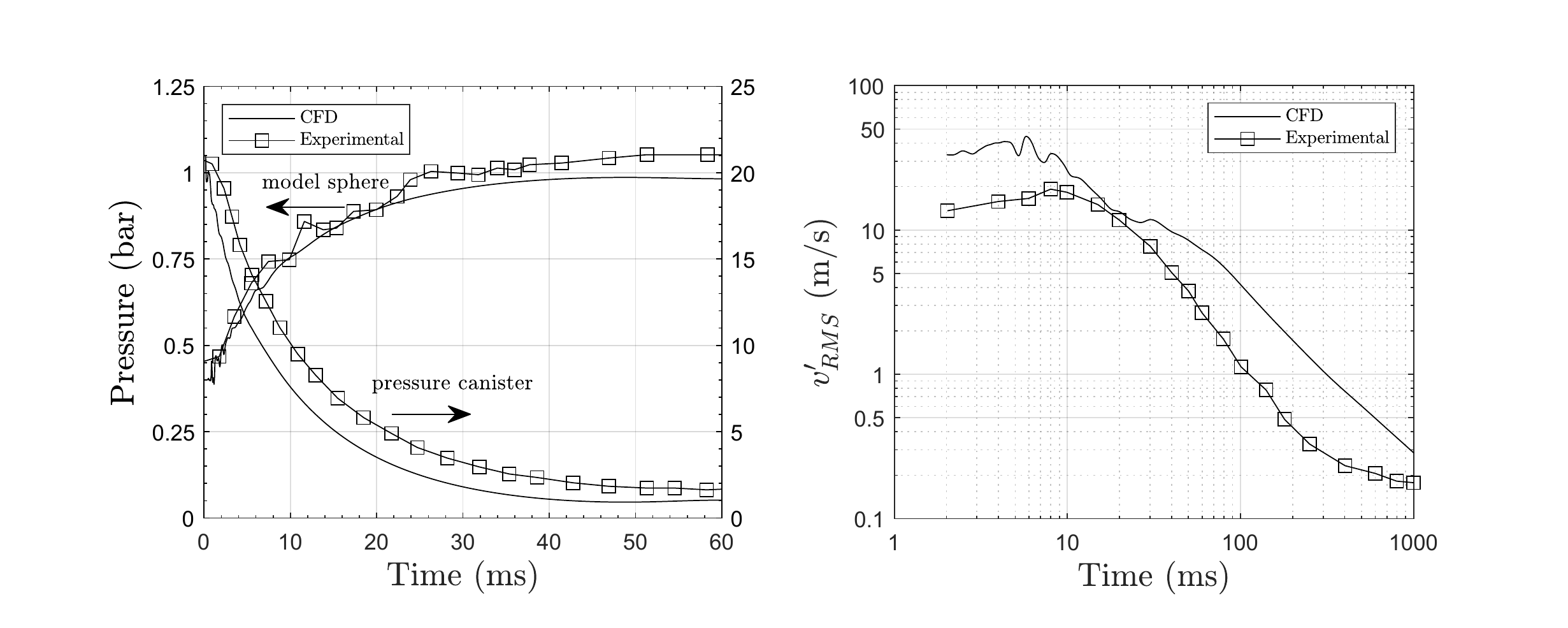}}
    
    \subfigure[(Air-dust flow) Comparison of the temporal trends of the pressure inside the sphere and canister (left) and evolution of the RMS velocity fluctuations in the sphere (right) between a study employing a commercial CFD code (Portarapillo et al. \cite{portarapillo2020cfd}) and CFD results obtained with OpenFOAM. \label{Subfigure:Validation_Portarapillo}]{\includegraphics[width=1\textwidth]{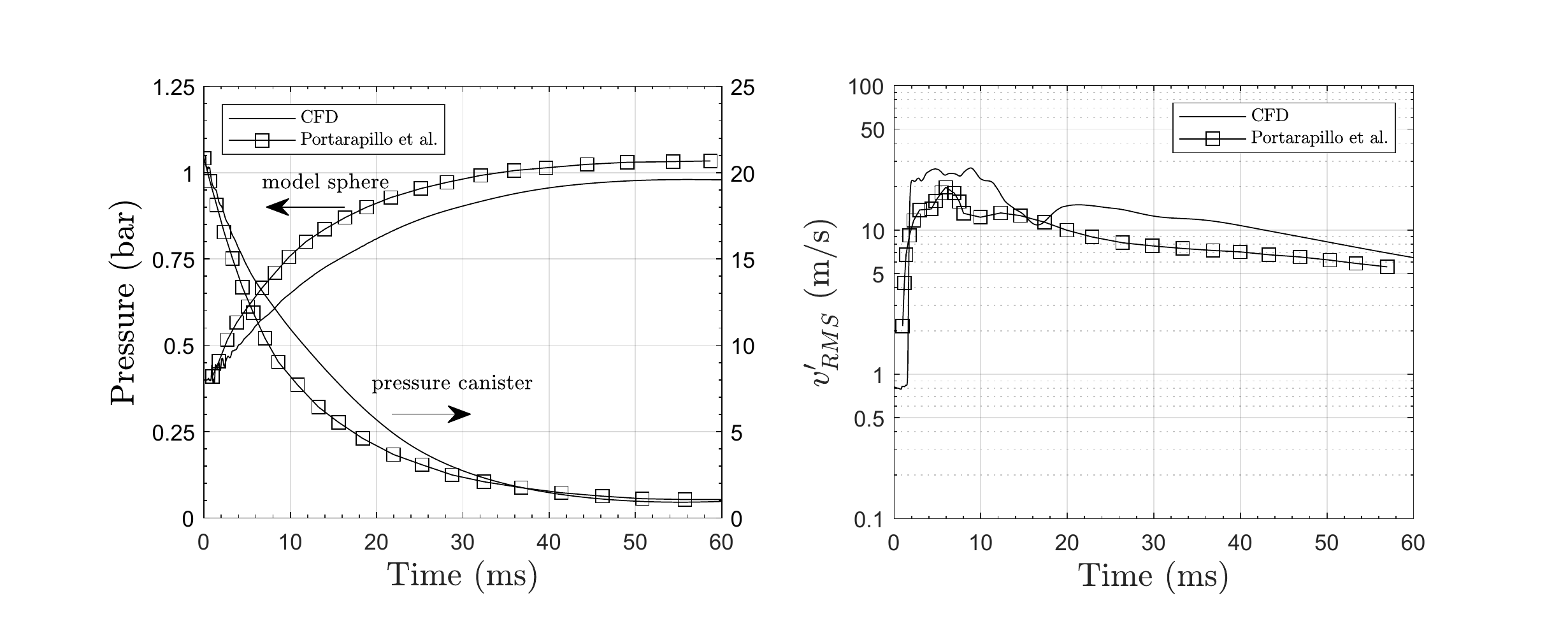}}
    
    \caption{Validation of the CFD model}
    \label{Figure:Validation_CFD_model}
       
\end{figure}

As observed in Fig. \ref{Subfigure:Validation_Dahoe}, the trends of the pressure profiles are well captured as both the sphere and canister approximate asymptotically to the vicinity of 1 bar after $t=40 \text{ms}$. The latter suggests that the flow coming into the sphere, in this case being air only, is limited to this time. A slight underestimation of the pressure at the end of the ignition delay time is observed, with a relative error $\sim 3\%$. This can be explained due to the fact that the transient dispersion in the experiments is not isothermal, and the outlet valve is closed before the pressure and temperature reach equilibrium. In addition, the CFD profile of the pressure discharge in the canister is steeper than in the experimental measurements. This difference can be attributed to the acting mechanism of the outlet valve. As described in the UNE 14034 standard, the pressure release is more pronounced for the pneumatically-activated valves than for the fast-acting valves with blasting caps. The former is commonly utilized in the experimental study, while the latter resembles more to the profile obtained with the CFD model.\\

Similarly, the RMS velocity fluctuations were computed in a simulation run extended up to 1s, where both profiles exhibit a similar trend. First, as previously suggested by Dahoe et al. \cite{dahoe2001transient}, there is a period of turbulence build-up in which the baroclinc effect is dominant over other mechanisms of turbulence production. Particularly, this effect promotes turbulence production during an initial period of 10 ms, same in which the pressure in the canister is reduced about a 65\%. This condition takes the flow integration time step to reach values as low as $10^{-7}$ s. As the flow continues entering into the 20L vessel, the strength of the baroclinic effect decays significantly, such that mechanisms of turbulence production associated to wall friction and shear stresses are not able to overcome this decay, hence leading to a general decay of the RMS velocity fluctuations in the sphere.\\

Second, the CFD model reports a moderate overestimation of the RMS velocity fluctuations during the whole run. This can be explained due to the fact that the geometry considered in the model does not account for the pyrotechnic ignitors at the center of the sphere. These cylindrical bodies represent an obstacle to the fluid flow in the region where the $v'_{\text{RMS}}$ was sampled. This increases the turbulence intensity at that zone, therefore leading to a larger values of the velocity fluctuations. Most important, the CFD model predicts an exponential-like decay of turbulence once past the 60 ms, as earlier observed by other authors \cite{dahoe2001transient,MURILLO201854}. Finally, in the current CFD model the temperature inside the canister at the end of the dispersion process is 123 K, only a 1.8\% deviation from the estimation if the discharge is considered as an isentropic process.\\

Later, the model was compared to the numerical results of Portarapillo et al. \cite{portarapillo2020cfd}. In their study, the dispersion of a niacin dust was simulated by using the commercial CFD code ANSYS Fluent\textsuperscript{\textregistered}. The properties of the dust sample were imitated, namely $\rho_{p}=$ 1470 $\text{kg}/\text{m}^3$ and fixed $d_{p}=$ 41.4 $\mu\text{m}$ for a case with a dust concentration of 250 $\text{g}/\text{m}^3$. The profiles of the evolution of the pressure and TKE are shown in Fig. \ref{Subfigure:Validation_Portarapillo}. As appreciated, the pressure inside the sphere is somewhat underestimated, with the profile of the commercial CFD code being more approximate to that obtained with OpenFOAM in the case with air only. This difference can be attributed to the fact that the injection method in both CFD codes are different. The former employs a surface injection controlled by particle mass flow rate, while the latter places all the computational parcels into the frustum section of the canister at stagnant conditions. The last-mentioned injection method was employed with the aim of elucidating what happens in the real experimental setup, as the dust samples are initially quiescent upon the activation of the outlet valve, which thereafter allows the carrier phase to drive the particles. Another difference is that the present study emulates a more realistic bottle-like shape of the canister and keeps the curvature of the connection pipe, whereas the model simulated with the commercial CFD code considers a spherical canister and a straight vertical pipe.\\

In addition, the profiles of the pressure discharge in the canister match reasonably well, except for the period between 5 and 25 ms, where the most of the particle flow is assumed to enter into the sphere. A similar discrepancy is observed for the pressure increase in the sphere, suggesting that the presence of particles in the canister attenuate both the rate of pressure rise in the sphere and the rate of pressure discharge in the canister. Furthermore, as no experimental data on the pressure profiles obtained in the air-dust flow is available, the temporal trends of the pressure and RMS velocity fluctuations obtained with OpenFOAM are considered of satisfactory performance due to the final pressures and $v'_{\text{RMS}}$ at the end of the ignition delay time approach to the vicinity of 1 bar and 6 $\text{m}/\text{s}$, respectively.\\




\subsection{Analysis of the flow pattern}

\begin{figure}
    \centering
    \includegraphics[width=0.85\textwidth]{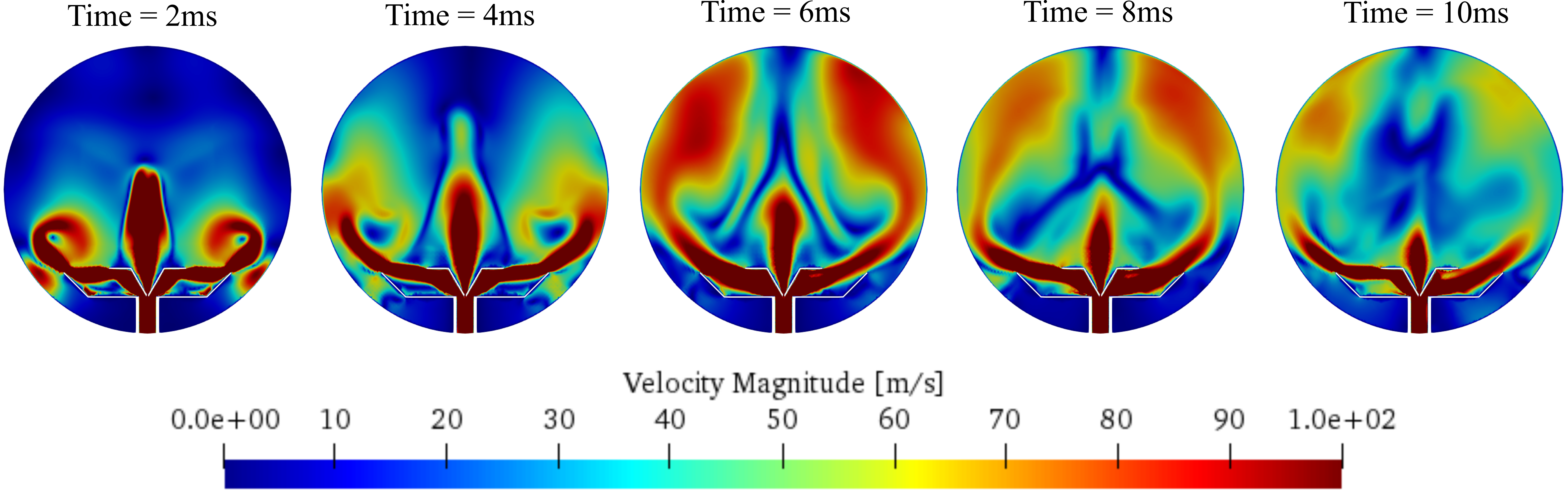}
    \caption{Snapshots of the velocity contours during the initial 10 ms of the air blast.}
    \label{Fig:velocity_contours}
\end{figure}

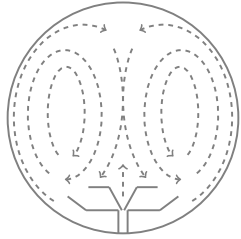
\begin{figure}
    \centering
    \adjustbox{width=0.20\textwidth}{
     \begin{tikzpicture}[very thick]
        \draw (0,0) circle [radius=2.5cm];
        \draw (-0.1,-2.5) -- (-0.1,-2.0);
        \draw (0.1,-2.5) -- (0.1,-2.0); %
        \draw (-0.1,-2.0) -- (-0.8,-2.0);
        \draw (0.1,-2.0) -- (0.8,-2.0);
        \draw (-0.8,-2.0) -- (-1.2,-1.7);
        \draw (0.8,-2.0) -- (1.2,-1.7);
        \draw (-0.025,-2.0) -- (-0.3,-1.5);
        \draw (0.025,-2.0) -- (0.3,-1.5);
        \draw (-0.3,-1.5) -- (-0.75,-1.5);
        \draw (0.3,-1.5) -- (0.75,-1.5);
        
        \draw [->=stealth, dashed] (-1.5,-1.8) arc (245:70:1.5cm and 2cm);
        \draw [->=stealth, dashed] (-0.2,1.5) .. controls (0.1,0.8) and (0.1,-0.5)  .. (-0.5,-1.3);
        \draw [->=stealth, dashed] (-1.6,-1.2) arc (245:-90:0.8cm and 1.4cm);
        \draw [->=stealth, dashed] (-1.4,-0.8) arc (245:-70:0.4cm and 1cm);

        \draw [->=stealth, dashed] (1.5,-1.8) arc (-65:110:1.5cm and 2cm);
        \draw [->=stealth, dashed] (0.2,1.5) .. controls (-0.1,0.8) and (-0.1,-0.5)  .. (0.5,-1.3);
        \draw [->=stealth, dashed] (1.6,-1.2) arc (-65:270:0.8cm and 1.4cm);
        \draw [->=stealth, dashed] (1.4,-0.8) arc (-65:250:0.4cm and 1cm);
        
        \draw [->=stealth, dashed] (0, -1.7) -- (0,-1);
        
    \end{tikzpicture}
    }
    \caption{Flow pattern in the 20L sphere.}
    \label{Figure:Flow_pattern}

\end{figure}

As previously stated, the period of turbulence build-up is limited to the first 10 ms of the dispersion process. During this time the incoming flow to sphere reaches a sonic condition, with maximum velocities circa 350 $\text{m}/\text{s}$. Fig. \ref{Fig:velocity_contours} shows the velocity contours at a cross-sectional plane of the sphere during this time. As observed, three jets are expelled out from the nozzle holes resembling a trident-like shape. The lateral jets are directed towards the walls, while the central jet is directed up to the top of the vessel. As the flow evolves, the lateral jets continue adjacent to the walls until they collide at the top of the sphere and start to descend. The descending flow is then curbed and deflected by the central jet into two symmetrical opposite directions, hence creating the two re-circulation regions shown in Fig. \ref{Figure:Flow_pattern}. \\

\begin{figure}
    \centering
    \includegraphics[width=0.3\textwidth]{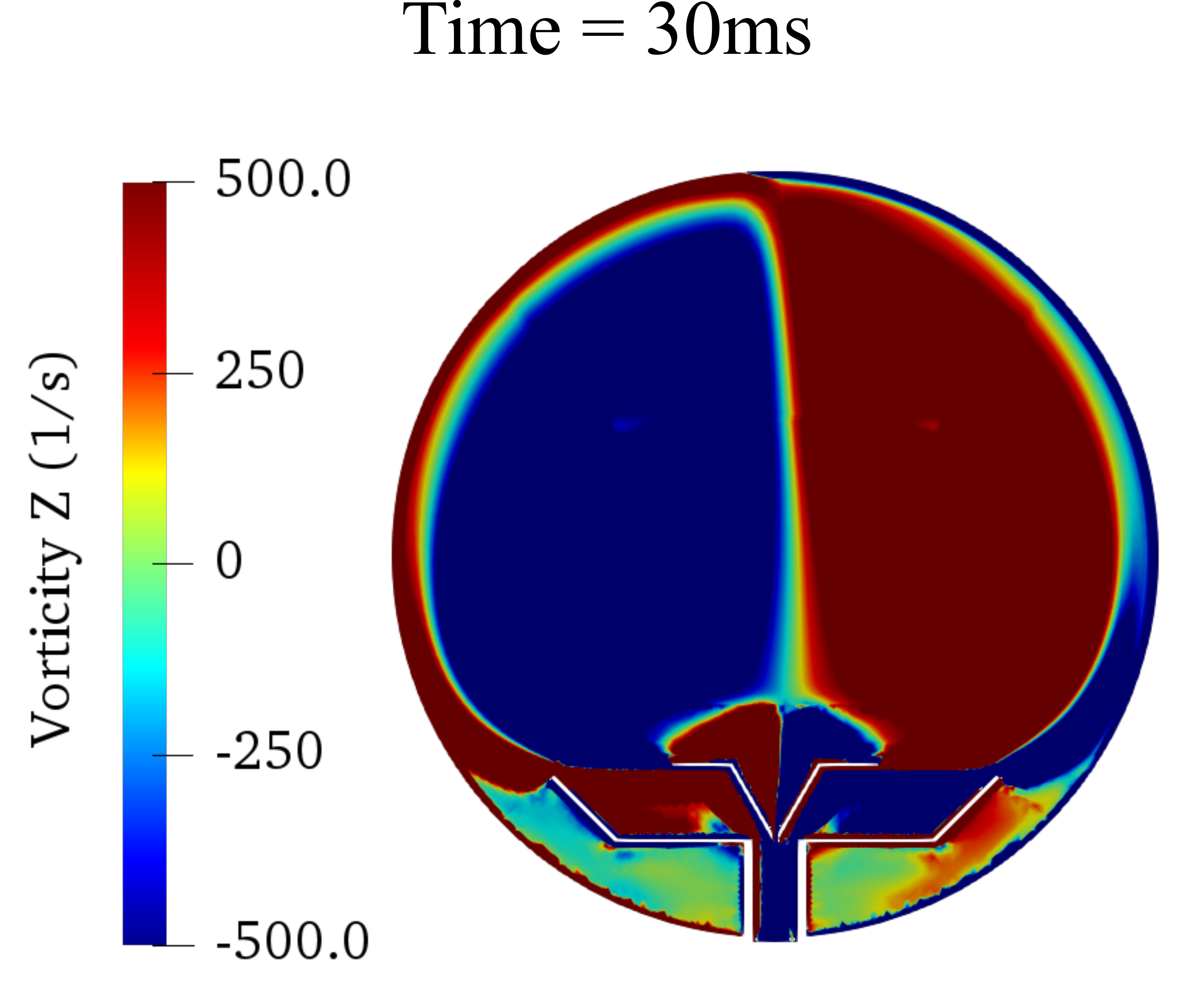}
    \caption{Snapshot of the z-component of the vorticity field at $t=30 \text{ms}$.}
    \label{Figure:Vorticity}
\end{figure}

The visualization of this pattern is supported by the experiments of Du et al. \cite{du2015visualization}, who used a high-speed camera and image processing techniques to study the dispersion process in a transparent 20L vessel. In their study, they considered carbonaceous dust samples, finding that the presence of these two re-circulation regions lead to a spatial non-homogeneous distribution of the dust cloud. Similarly, the pattern was also observed in the CFD studies of Di Benedetto et al. \cite{DIBENEDETTO2013cfd} and Di Sarli et al. \cite{di2014cfd}, in which the effect of the re-circulation regions on the distribution of the dust particles inside the sphere were highlighted. They suggest that these vortices act as dead volumes for the gas-solid flow, minimizing the dust concentration at the core regions. Moreover, in Fig. \ref{Figure:Vorticity} the contours of the z-component of the vorticity field illustrate the vortex direction of rotation at a later time during the dispersion process. Here the vorticity is positive for the counter-clockwise rotating vortex, while it is negative for the vortex rotating in the clockwise direction. This map also evidences the existence of strong shear layers in zones with sharp red-to-blue (and vice-versa) color transitions. Besides taking place at the vertical axis of symmetry, these are especially appreciable in the zones between the upper and lower plates of the nozzle. The latter are further extended to the tips of the lower plates until continuing adjacent to the walls of the sphere. In general, this two-vortex pattern prevails until the end of the ignition delay time and has a direct impact on the mixing of the dust cloud.\\

\begin{figure}
    \centering
    \includegraphics[width=0.85\textwidth]{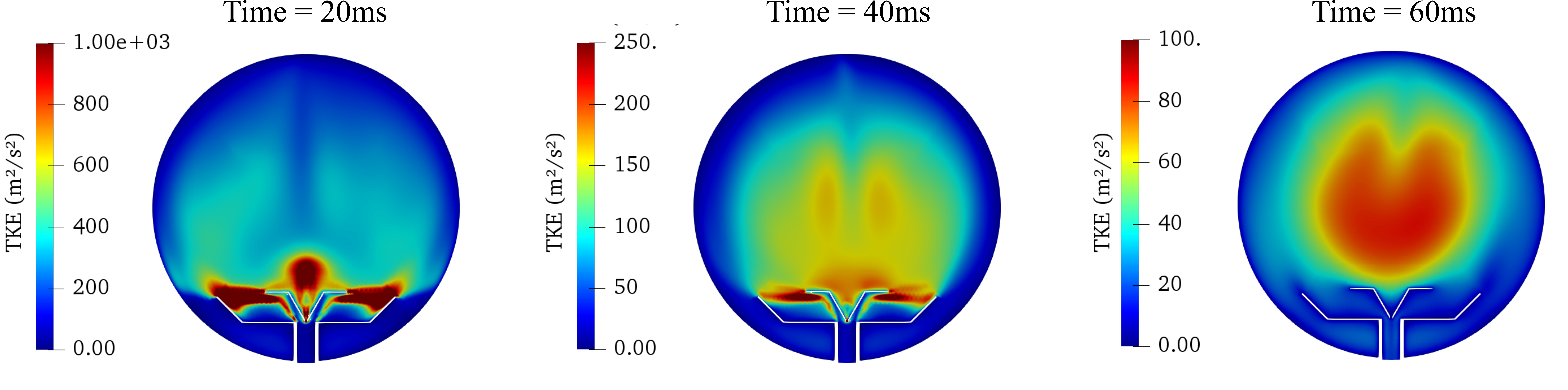}
    \caption{Snapshots of the TKE contours at selected times during the air blast.}
    \label{Figure:TKE_air}
\end{figure}

Likewise, Fig. \ref{Figure:TKE_air} shows the TKE contours at selected times during the dispersion process. First, during all times the TKE map is symmetrical with regard to the vertical axis passing through the center of the sphere. Next, it is observed that at $t=$ 20 $\text{ms}$, the zones with higher TKE are between the lower and upper plates of the nozzle and at the exit of the central jet. The latter agrees with the vorticity contour, indicating a high swirling flow taking place in these regions. Then, at $t=$ 40 $\text{ms}$ the regions of higher TKE are contained within the cores of the two-vortex structure described earlier.
Lastly, by the end of the ignition delay time, $t_{\text{d}}$, the higher TKE is still concentrated at the center of the sphere, suggesting that the flow is relatively more turbulent at this region than at the near-wall regions in the sphere. The latter will have an impact on the turbulence-chemistry interaction for an extended study considering the reactive scenario.\\

\begin{figure}
    \centering
    \includegraphics[width=\textwidth]{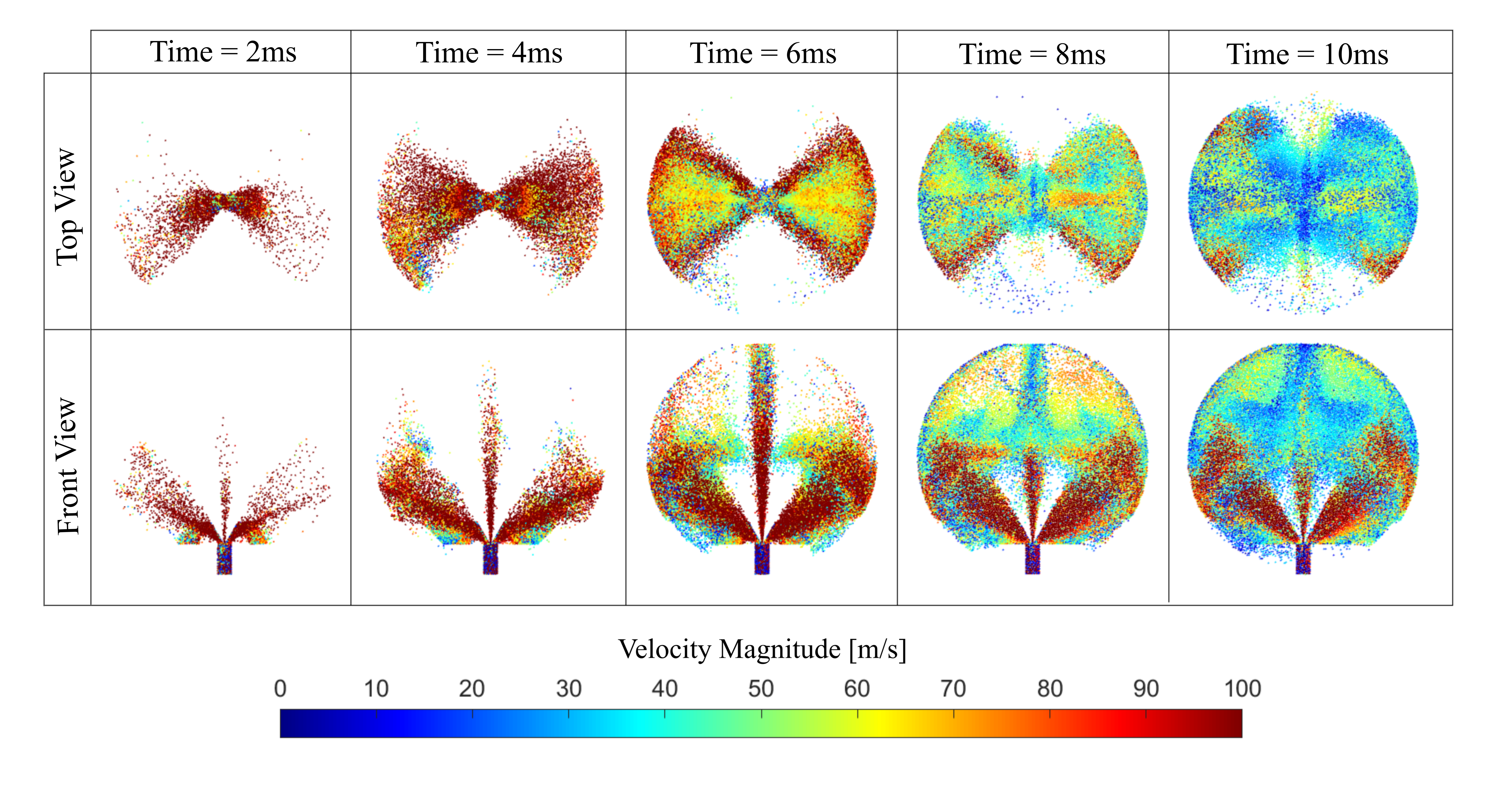}
    
    \caption{Snapshots of the particle tracks colored by velocity magnitude during the initial 10ms of dispersion of the air-dust mixture. The top and front views of the sphere are portrayed in the first and second rows, respectively.}
    \label{Fig:dust_dispersion}
\end{figure}

Similarly, Fig. \ref{Fig:dust_dispersion} shows instantaneous snapshots of the evolution of the particle dispersion during the initial 10 ms of the dispersion of the air-dust mixture. Here, the effect of the rebound nozzle is clearly appreciated at the first two frames. The particles expelled out from the central holes are directed up to the top of the sphere while the particles coming out from the lateral holes are ejected towards the upper plates of the nozzle. The latter particles bounce back and are then reflected by the lower plates, following the same jet behavior of the gas phase. Furthermore, the top view of the sphere evidences that the lateral particle jets are also spread into a symmetric cross-shaped pattern with regard to the horizontal plane. \\

As the time elapses, the particles continue mixing with the flow while those hitting the walls are reflected with elastic conditions. Furthermore, during the initial 10 ms of dispersion, approximately a 50\% of the total particle mass fraction enters into the sphere. This is attributed to the fact that flow is particularly vigorous during this time.\\



\subsection{Effect of particle size on the turbulence of the particle laden flow}

Next, with the aim of studying the effect of particle size on the turbulence of the carrier phase, four simulations with varying particle diameter were performed. The particles are assumed to have a density $\rho_{\text{p}}=$ 1400 $\text{kg}/\text{m}^3$, which corresponds to a classic carbonaceous or woody dust sample. A concentration of 250 $\text{g}/\text{m}^3$ is considered for all cases, with the following fixed particle diameters: 10, 50, 100 and 200 $\mu\text{m}$. These values are typical for comminuted materials, and lie in the most explosive size range of particulate solids processed in the power, mining and pharmaceutical industries. To obtain statistically significant results, in all cases the total mass of the dust sample was equally distributed among 1M computational parcels.\\

\begin{figure}
    \centering

    \subfigure[Comparison of the temporal trends of the TKE in the 20L sphere between the unladen and particle-laden flows with varying particle diameter (left), and change of the turbulence intensity between the unladen and particle-laden flows (right). \label{Subfigure:Turbulence_modulation}]{\includegraphics[width=\textwidth]{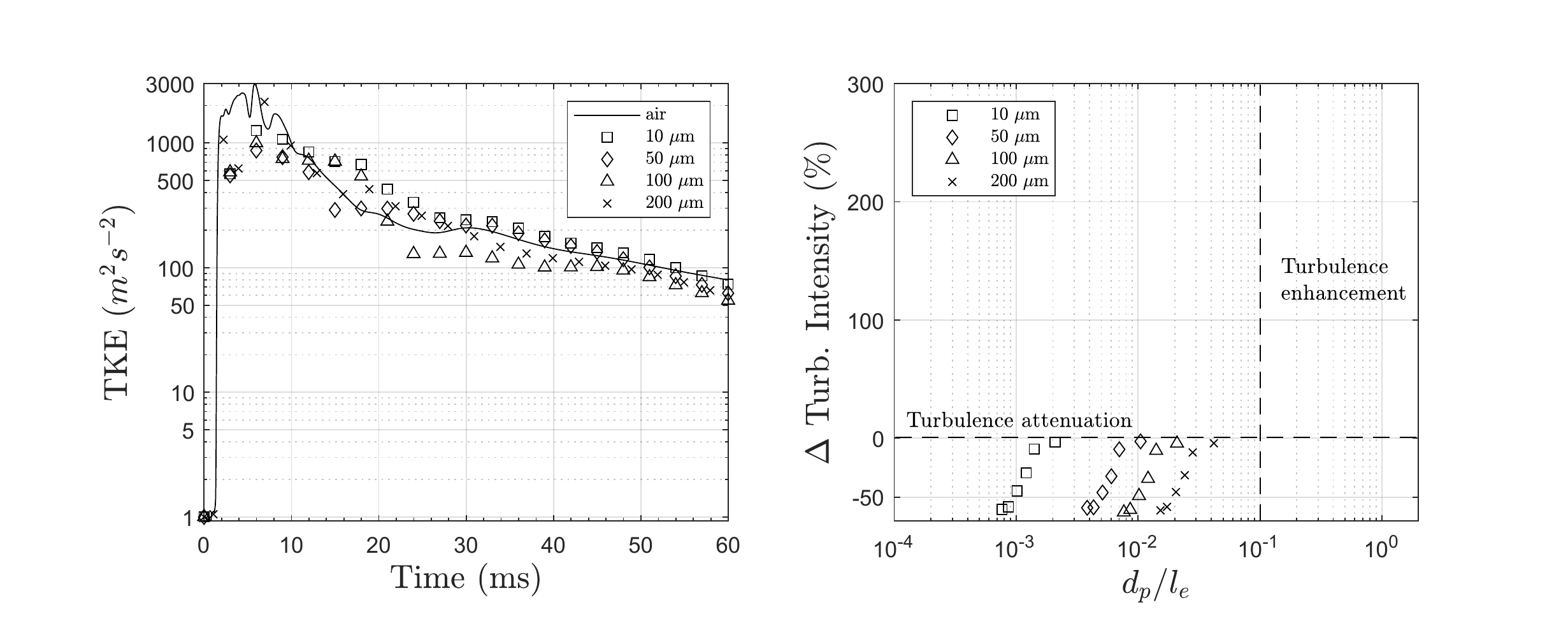}}
    
    \subfigure[Comparison of the temporal evolution of the particle Reynolds number (left) and Stokes number (right) between the particle laden flows with varying particle diameter. \label{Subfigure:Dimensionless_numbers}]{\includegraphics[width=\textwidth]{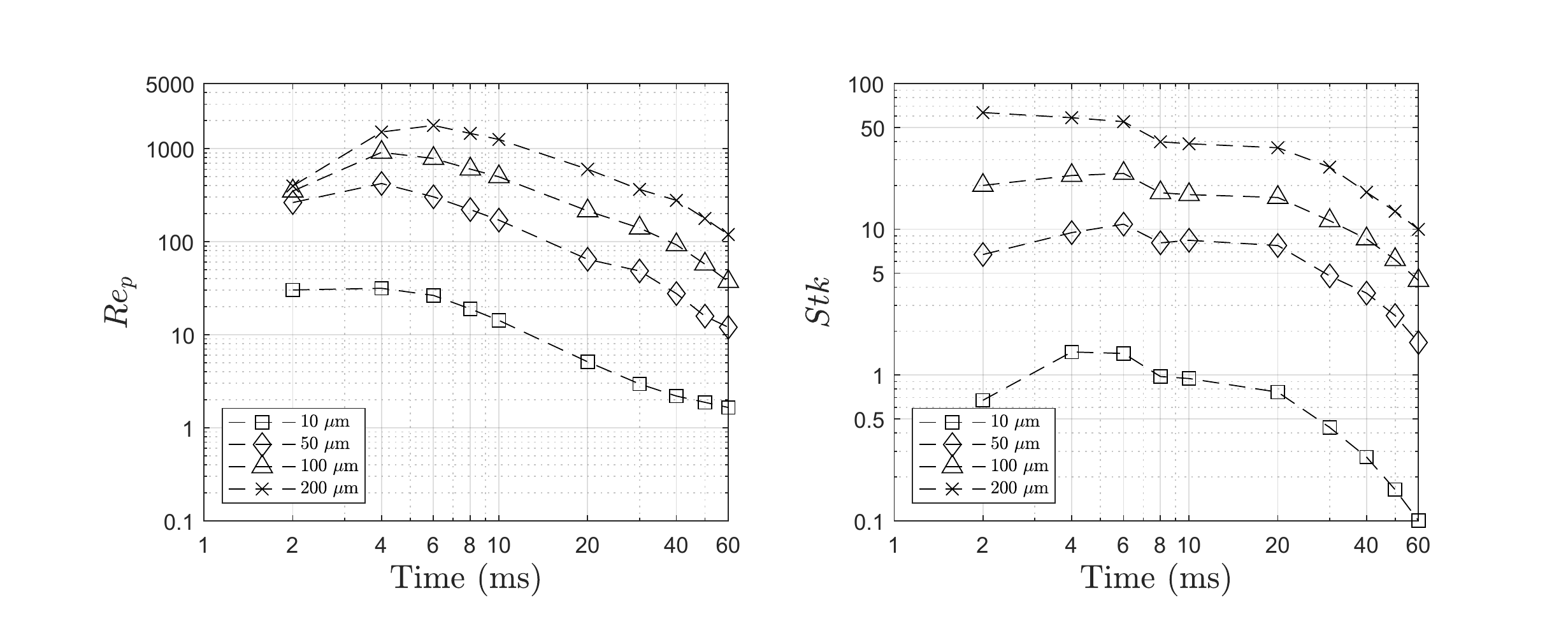}}

    \caption{Effect of the presence of a dispersed phase on the turbulence of the dust-free case and dynamics of the dust particles.}
    \label{Figure:TKE_particles}
\end{figure}

Fig. \ref{Subfigure:Turbulence_modulation} shows a comparison of the temporal trends of the TKE between the dust-free case and the cases with varying particle diameters. First, it is appreciated that the periods of turbulence build-up and decay are still in agreement with regard to that of the dust-free simulation. During the initial 10 ms of the dispersion process, the TKE of all the two-phase flows is lower compared to that of the single-phase flow. The latter can be explained by the work of Balachandar and Eaton \cite{balachandar2010turbulent}, who propose three mechanisms of turbulence reduction for dilute suspensions (the volume fraction is $\phi \sim 10^{-4}$ in this study): (a) enhanced inertia of the particle-laden flow, (b) increased dissipation due to particle drag, and (c) enhanced kinematic viscosity, $\nu_{\text{eff}}$, of the particle-laden fluid. These effects become relevant when the particle scales are comparable to the Kolmogorov scales. In this study, particles are in the range of tens of microns, $\mathcal{O}\left(d_p\right)=10^{-5}$, while the Kolmogorov time scales calculated during the period of turbulence build-up, $\mathcal{O}\left(\eta\right)=10^{-5}$, therefore suggesting important local flow distortion around the particles.\\

Contrarily, turbulence can be enhanced due to wake dynamics and self induced vortex shedding around the particles, as it happens for a short time period between 15 and 20 ms. During this time, the TKE is larger than that of the dust-free flow, suggesting that particles promote unsteady wakes that are not present in the unladen flow. Next, for the time period past the initial 20 ms, the particles with $d_{\text{p}}\geq$ 100 $\mu\text{m}$ decrease the TKE, while particles with $d_{\text{p}}\leq$ 50 $\mu\text{m}$ exhibit a slight TKE increase that is further reduced up to reaching similar values than those obtained with the injection of larger particles. In general, during the period of turbulence decay, the presence of particles attenuates turbulence. This can be explained by the turbulence modulation phenomenon. \\

As suggested by Crowe et al. \cite{crowe2011multiphase}, the turbulence modulation can be qualitatively classified in a map divided into two regions. These zones are distinguished by a criterion based on surface effects, namely the ratio of the particle diameter $d_{\text{p}}$ to the dissipation length scale $l_{e}$. This is particularly useful to estimate up to which threshold value of $d_{\text{p}}$, the  presence of a dispersed phase in the flow can either enhance or attenuate the turbulence relative to the turbulence of the carrier phase. According to Crowe's, this ratio is $ d_{\text{p}}/l_{e}\sim 0.1 $. This map is depicted similarly in Fig. \ref{Subfigure:Turbulence_modulation} right, portraying the regions of turbulence modulation originally shown in Crowe's work. It is observed that the presence of all particles ranging from 10 to 200 $\mu\text{m}$ attenuate the turbulence intensity during the period of turbulence decay, as the $d_{\text{p}}/l_{e}<0.1$ for the four cases above mentioned. This reduction will have an impact on the rates of heat transfer and chemical reactions for the dust explosion test \cite{crowe2011multiphase}.\\

In addition, from the works of Ferrante and Elghobashi \cite{ferrante2003physical} and Kussin and Sommerfeld \cite{kussin2002experimental} it is suggested that there are other factors that appear to contribute to turbulence modulation due to the presence of particles. These include (a) inertial effects: particle Reynolds number, and  (b) response effects: particle response time, or Stokes number. Fig. \ref{Subfigure:Dimensionless_numbers} shows the evolution of these dimensionless numbers for the four particle-laden flows analyzed in this section. As appreciated, all profiles of the particle Reynolds number ($Re_{\text{p}}$) follow a similar trend during the 60 ms of dispersion. All peak values are reached during the initial 10 ms, which corresponds to that of the period of turbulence build-up. Then all profiles exhibit a smooth decay until the end of the ignition delay time. However, because of the length scales at which these different mechanisms vary, the suspended particles can simultaneously augment and suppress over a different range of scale, such that the effective modulation depends on the strength of the different mechanisms.\\

Moreover, the first map suggest that the dynamics of the dust particles are far from the so-called Stokes regime, in which particle inertia is small compared to that of the gas-phase. Conversely, all the particles considered in this study happen to drift from the fluid streamlines due to a high inertial force ($Re_{\text{p}}>1$), except those having $d_{\text{p}} =$ 10 $\mu\text{m}$. An analogy can be made to interpret the second map, as a \textit{Stk} $<1$ indicates that the dust particles are able to adapt to the changes of the flow field, while for \textit{Stk} $>1$ the particle response times are larger than the characteristic fluid time scales suggesting that the interaction between the dynamics of the two phases is minimal.\\

In short review, all particles exceeding $d_{\text{p}}\geq 50\mu\text{m}$ will not follow the flow pattern, as both $Re_{\text{p}}$ and \textit{Stk} numbers indicate that these particles are far from reaching equilibrium with the inertia of the gas-phase. This fact will have an impact on the distribution of the dust cloud. \\

\subsection{Effect of particle size on the distribution of the dust cloud}

\begin{figure}
    \centering
    \includegraphics[width=\textwidth]{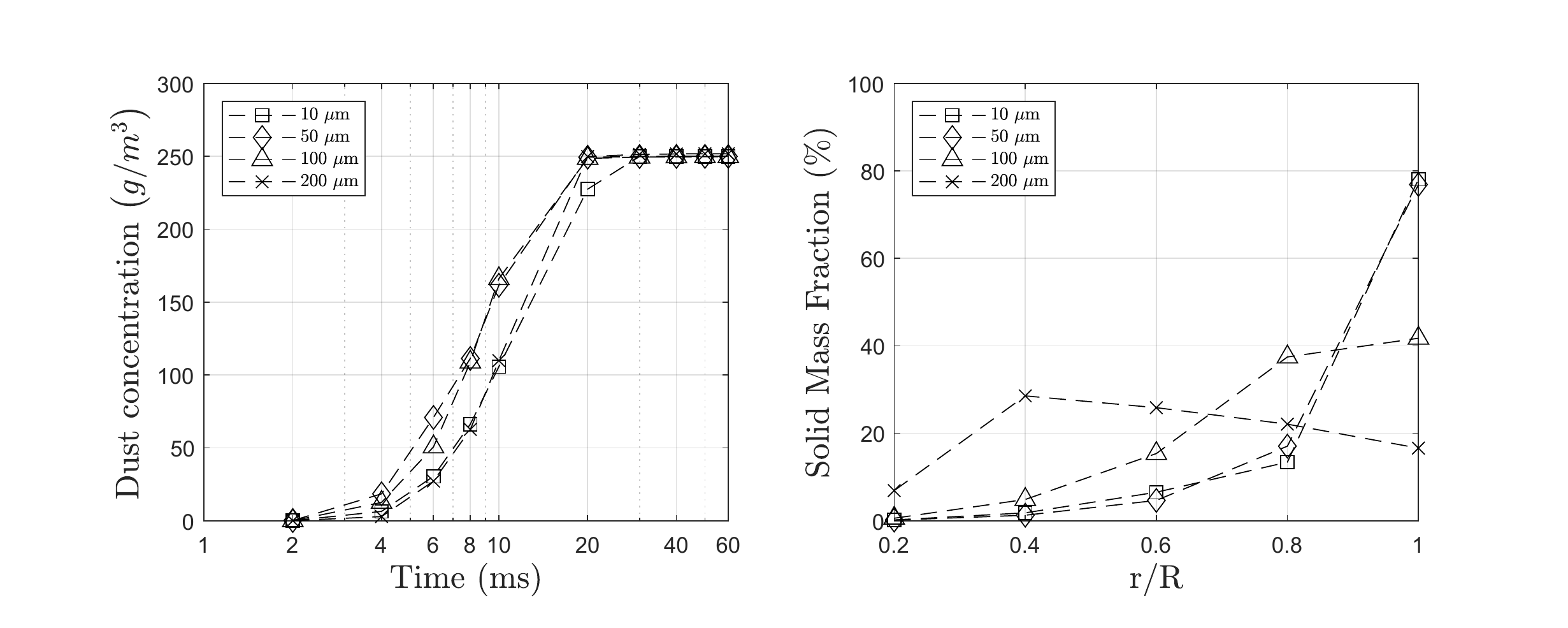}
    \caption{Temporal evolution of the dust concentration attained in the 20L sphere for various particle sizes (left), and distribution of the dust particles versus radial position for various particle sizes at $t=60\text{ms}$ (right).}
    \label{Figure:Dust_distribution}
\end{figure}

The temporal evolution of the dust concentration attained in the 20L sphere is plotted in Fig. \ref{Figure:Dust_distribution}. In all cases, the dust filling in the vessel follows a linear trending, similar to the observations of Di Benedetto et al. \cite{DIBENEDETTO2013cfd}. Moreover, the time for which the 
total of the dust sample is injected into the chamber seems to be size independent, as all cases reach the asymptotic value of the nominal dust concentration for times circa 20 - 30 ms. This finding suggests that all the dust particles have about half of the ignition delay time to mix with the coherent flow structures. However, as described earlier, in any case the interaction of the dynamics between solid and gas phases is optimal. The dominant vortex structures in the flow field is a mechanism of reduced particle mixing due to turbulence. The latter is associated to the nozzle design and spherical geometry of the 20L vessel. Although some authors \cite{MURILLO201854,yao2020analysis} have proposed other nozzles to improve the particle mixing, no conclusive data has been obtained yet.\\

Furthermore, to evaluate the dust distribution inside the 20L sphere at the end of the ignition delay time, particles were measured as if contained in concentric spherical zones inside the vessel. In particular, five radial regions were considered; the first one being a perfect sphere at $\text{r}/\text{R}=0.2$, while the rest of the regions were drafted as spherical shells of the subsequent corresponding radii. As also noted in Fig. \ref{Figure:Dust_distribution}, the cases with the two smallest particle size ($d_{\text{p}}\leq$ 50 $\mu\text{m}$) exhibit a very similar profile, with a preferential concentration of the dust particles in the outermost shell. The mass of particles contained in this shell represents about an $80\%$ of the total dust sample and is concentrated in the near-wall region. This may have a significant impact on the ignition of the air-dust mixture, as many particles lie far from the ignition zone by the moment in which the ignitors are activated. Then, it is appreciated that as the particle size increases, the dust distribution in the outermost shell decreases up to $20\%$. This is attributed to the increased inertia of the largest particles, as both $Re_{\text{p}}$ and \textit{Stk} numbers are about one order of magnitude larger compared to the smallest two particle sizes. In this case, the interaction effects with the walls play a major role for determining the final positions of the particles. \\

The particles with the two largest diameters carry more energy and momentum, exhibiting a ballistic behavior. Moreover this study considers elastic coefficients of restitution, therefore particles hitting the walls retain its normal and tangential momentum after the rebound. This implies that particles having diameters of 100 and 200 $\mu\text{m}$ will enter into the re-circulation regions with a negligible effect of the surrounding fluid on the particle trajectories. In consequence, the distribution of the dust cloud is more uniform when larger particles are injected. Previous work carried out by Di Sarli et al. with their 3D CFD model \cite{DiSarli2013727}, has also shown that dust paths differed from those of the fluid flow when dust diameter was increased. In that work, although a material density of around 46\% higher was considered, the same outcome was presented. Smaller particles can follow the flow pattern to a higher degree, then driven by the two recirculation zones tend to be more concentrated at the vessel walls. In contrast, larger diameter dust presents a high inertial force which makes them almost independent of the dynamics of the gas-phase, and thus, the interaction effects with the walls play a major role in their trajectories. As a result, particle size seems to be a key characteristic for dust dispersion over the material density.\\

\begin{figure}
    \centering
    
    \begin{tabular}{c|c}
    \subfigure[Front view \label{Subfigure:ParticlePositions_XY}]{\includegraphics[width=0.45\textwidth]{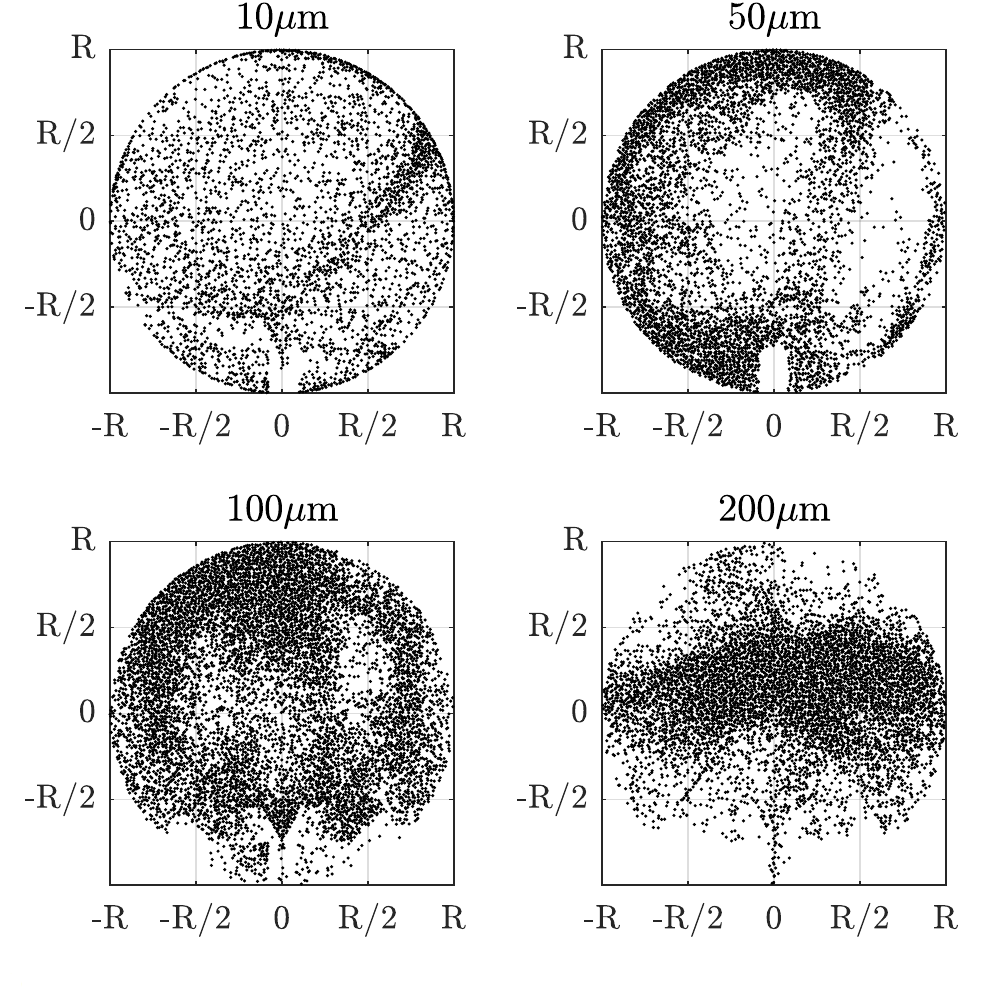}}     &   \subfigure[Top view \label{Subfigure:ParticlePositions_XZ}]{\includegraphics[width=0.45\textwidth]{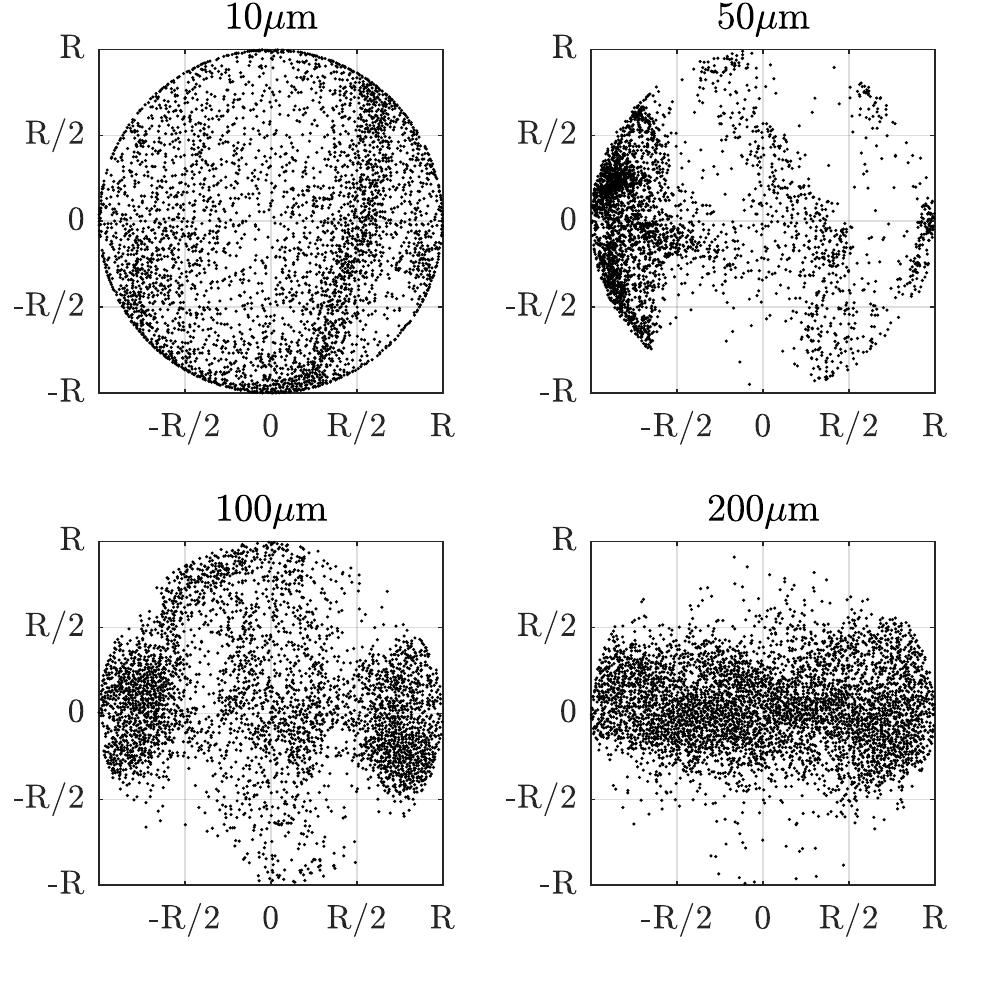}}

    \end{tabular}
   
    \caption{Distribution of the dust cloud sampled at a cross-sectional plane coincident to the: xy-plane or front view (left), and xz-plane or top view (right) of the sphere at $t=60 \text{ms}$.}
    \label{Figure:Particle_positions}
       
\end{figure}

To show the positions of the particles at the end of the ignition delay time for the isolated cases of $d_{\text{p}}$ considered in this study, Fig. \ref{Figure:Particle_positions} shows a graphical representation of the dust cloud. Here, the depicted particles correspond to those sampled at a transverse plane coincident to the xy-plane (or front view of the sphere) and to the xz-plane (or top view of the sphere).\\ 

These pictures agree with the aforementioned observations, as for the case with $d_{\text{p}}=$ 10 $\mu\text{m}$, there is a thin layer of particles adjacent to the walls and a little concentration at the center of the sphere. This distribution is similar to the case containing particles with $d_{\text{p}}=$ 50 $\mu\text{m}$, where a bigger amount of particles in the near wall region is appreciated. In addition, for this case it is noted that there is a minimal dust concentration in the zones occupied by the re-circulation regions. Therefore, the two vortices are dead volumes for the particles. Next, for the case with $d_{\text{p}}=$ 100 $\mu\text{m}$, more particles enter into the re-circulation regions, such that the dust concentration at the center of the sphere is increased. Finally, for the case containing particles with $d_{\text{p}}=$ 200 $\mu\text{m}$, a dense cloud at the center is observed, which in fact confirms that the peak of the solid mass fraction occurs at $\text{r}/\text{R}=0.4$. \\

To end, the present study points out an already discussed issue regarding the experimental dispersion nozzle. The standard rebound nozzle has been shown incapable of producing a homogeneous dust cloud, especially for larger dust particle sizes, as highlighted by Di Sarli et al. \cite{DiSarli2013727}.  Alternatively, using the perforated annular nozzle a better-mixed dust/air cloud can be generated but failing in the injection of the whole dust mass contained within the canister \cite{DISARLI2015204}. The aforementioned problems lead to a dust concentration completely different from the nominal value (under some conditions/locations) which may provide spurious results when measuring the explosion and flammability parameters considering the standard procedure.\\

\section{Conclusions}

Dust dispersion process in the standardized 20L spherical vessel was simulated by employing the open-source CFD code OpenFOAM. The model was validated against experimental measurements of the dust-free flow and with numerical results of air-dust flow obtained with a commercial CFD code. The model agrees reasonably well with both studies since it has shown capable of capturing the most significant features of the transient flow, such as the periods of turbulence build-up and decay and the two-vortex structure that dominates the flow pattern. \\

In addition, the analysis of the effect of particle size (with $d_{\text{p}}$ equal to 10, 50, 100 and 200 $\mu\text{m}$) on the turbulence of the gas phase and on the distribution of the dust cloud suggests the following findings: a) First, it is observed that in all cases the presence of particles attenuates the turbulence intensity of the carrier phase, having a direct impact on the rates of heat transfer and chemical reactions for the subsequent dust explosions; b) Second, the homogeneity of the dust cloud obtained with the rebound nozzle is strongly influenced by the particle diameter, as the increased inertia associated with a particle size in the order of hundreds of microns $\mathcal{O}\left(d_p\right)=10^{-4}$ allows these particles to enter into the re-circulation regions, therefore increasing the number of particles that are close to the ignitors. Contrarily, the high dust concentration at the near wall region (or poor dust distribution) is more susceptible to occur when smaller particles $\mathcal{O}\left(d_p\right)=10^{-5}$ are injected. \\


Furthermore, it is shown that dust particles with $d_{\text{p}}\geq$ 50 $\mu\text{m}$ have a negligible interaction with the dynamics of the gas-phase, as $Re_{\text{p}}>$100 and \textit{Stk} $>$1 indicate that particle inertia is dominant over that of the surrounding fluid. As $d_{\text{p}}$ keeps increasing, the particles exhibit a more ballistic behavior, such that the collisions with the walls play a major role on determining the final positions of the particles.\\ 



 In summary, this multi-phase simulation conducted employing the open-source CFD code OpenFOAM has proven its full validity showing a great agreement with experimental measurements and other results produced with commercial CFD codes. Considering that, turbulence behavior and particle size effect have been thoroughly assessed. Moreover, the strongest advantage of the present code is its capacity to mimicking the real experiment, placing the solid particles at stagnant conditions in the frustum section of the canister while generating the particle-laden flow through a pressure difference. This allows to consider flow asymmetries induced by a realistic experimental setup. In this regard, it has been proved that the current experimental apparatus fails to provide a homogeneous dust cloud and a fully valid solution has not been presented yet. Like other CFD studies, this work encourages the development of a new dust dispersion method with the aim of improving the accuracy of dust explosion measurements in the 20L sphere. Future work in this geometry will include biomass as material, complete particle size distributions, different dust concentrations and consider different nozzle geometries, all of them under reacting conditions which imply the modeling of new complex processes such as the ignition mechanism, devolatilization process, heterogeneous combustion and flame propagation. These following steps will be made trying to contribute to the knowledge in powder science and technology while increasing the accuracy of the experimental techniques, since they determine the effective prevention and mitigation measures of dust explosions in the industrial field. \\





\section*{Declaration of Competing Interest}
The authors declare that they have no known competing financial interests or personal relationships that could have appeared to influence the work reported in this paper. \\

\section*{CRediT Authorship Contribution Statement 
}
\textbf{A. Islas:} Conceptualization, Formal analysis, Data curation, Methodology, Software, Validation, Investigation,  Resources, Writing - original draft, Writing - review \& editing, Visualization. \textbf{A. Rodríguez-Fernández:}  Methodology, Software, Validation, Investigation,  Resources, Writing - original draft, Writing - review \& editing. \textbf{E. Martínez-Pañeda:} Conceptualization, Writing - review \& editing, Funding acquisition. \textbf{C. Betegón:} Writing - review \& editing, Supervision, Project administration, Funding acquisition. \textbf{A. Pandal:} Conceptualization, Methodology, Software, Investigation, Resources, Writing - review \& editing, Supervision, Funding acquisition. \\

\section*{Acknowledgement}

Authors acknowledge that this work was partially funded by CDTI (Centro para el Desarrollo Tecnológico Industrial de España, IDI-20191151), Universidad de Oviedo and PHB WESERHÜTTE, S.A., under the project "FUO-047-20: Desarrollo de silo metálico de grandes dimensiones ante los condicionantes de explosividad de la biomasa". A. Islas acknowledges the support from the researh grant \#BP20-124 under the 2020 Severo Ochoa Pre(Doctoral) Program of the Principality of Asturias.\\




\bibliographystyle{unsrt_abbrv_custom}

\bibliography{cas-refs}

\begin{thebibliography}{10}

\bibitem{ogle2016dust}
R.~Ogle.
\newblock {\em Dust explosion dynamics}.
\newblock Butterworth-Heinemann, 2016.

\bibitem{SCHULZ201937}
D.~Schulz, N.~Schwindt, E.~Schmidt, R.~Jasevičius, and H.~Kruggel-Emden.
\newblock Investigation of the dust release from bulk material undergoing
  various mechanical processes using a coupled {DEM/CFD} approach.
\newblock {\em Powder Technology}, 355:37--56, 2019.

\bibitem{cheremisinoff2014dust}
N.~P. Cheremisinoff.
\newblock {\em Dust Explosion and Fire Prevention Handbook: A Guide to Good
  Industry Practices}.
\newblock John Wiley \& Sons, 2014.

\bibitem{eckhoff2003dust}
R.~K. Eckhoff.
\newblock {\em Dust explosions in the process industries: identification,
  assessment and control of dust hazards}.
\newblock Elsevier, 2003.

\bibitem{csb2006investigation}
{CSB, US}.
\newblock Investigation report, combustible dust hazard study.
\newblock {\em US Chemical Safety and Hazard Investigation Board}, 2006.

\bibitem{CLONEY201933}
C.~Cloney and J.~Snoeys.
\newblock Chapter three - dust explosions: A serious concern.
\newblock In P.~R. Amyotte and F.~I. Khan, editors, {\em Dust Explosions},
  volume~3 of {\em Methods in Chemical Process Safety}, pages 33--69. Elsevier,
  2019.

\bibitem{DSS_2020}
C.~Cloney.
\newblock {\em Combustible Dust Incident Report}, 2020.

\bibitem{ASTME1226}
ASTM International.
\newblock {\em ASTM E1226-19, Standard Test Method for Explosibility of Dust
  Clouds}, 2019.
\newblock West Conshohocken, PA, www.astm.org.

\bibitem{EN14034}
CEN, the European Committee for Standardization.
\newblock {\em EN 14034 - Determination of explosion characteristics of dust
  clouds}, 2011.

\bibitem{ISO6184-1}
International Standard Organisation, Geneva.
\newblock {\em ISO6184-1 Explosion protection systems - Part 1: Determination
  of explosion indices of combustible dusts in air}, 1985.

\bibitem{bartknecht1971brenngas}
W.~Bartknecht.
\newblock Brenngas und staubexplosionen forschungsbericht f45.
\newblock {\em Bundesinstitut Fur Arbeitsschutz (Bifa), Koblenz}, 1971.

\bibitem{bartknecht1980explosionen}
W.~Bartknecht.
\newblock {\em Explosionen: Ablauf und schutzmassnahmen}.
\newblock Springer, 1980.

\bibitem{DIBENEDETTO2007303}
A.~{Di Benedetto} and P.~Russo.
\newblock Thermo-kinetic modelling of dust explosions.
\newblock {\em Journal of Loss Prevention in the Process Industries},
  20(4):303--309, 2007.
\newblock Selected Papers Presented at the Sixth International Symposium on
  Hazards, Prevention and Mitigation of Industrial Explosions.

\bibitem{NFPA68}
National Fire Protection Association, 2013.
\newblock {\em Standard on Explosion Protection by Deflagration Venting, NFPA
  68}, 2018.

\bibitem{TASCON2011717}
A.~Tascón, A.~Ruiz, and P.~J. Aguado.
\newblock Dust explosions in vented silos: Simulations and comparisons with
  current standards.
\newblock {\em Powder Technology}, 208(3):717--724, 2011.

\bibitem{TORNO2020310}
S.~Torno, J.~Toraño, and I.~Álvarez Fernández.
\newblock Simultaneous evaluation of wind flow and dust emissions from conveyor
  belts using computational fluid dynamics {CFD} modelling and experimental
  measurements.
\newblock {\em Powder Technology}, 373:310--322, 2020.

\bibitem{THEIMER1973137}
O.~Theimer.
\newblock Cause and prevention of dust explosions in grain elevators and flour
  mills.
\newblock {\em Powder Technology}, 8(3):137--147, 1973.

\bibitem{BARTKNECHT1966}
W.~Bartknecht.
\newblock Gasexplosion in rohrstrecken.
\newblock {\em Bergfreiheit, Zeitschrift fiir den deutschen Bergbau}, 5, 1966.

\bibitem{siwek197720}
R.~Siwek.
\newblock 20-l laborapparatur f{\"u}r die bestimmung der
  explosionskenngr{\"o}{\ss}en brennbarer st{\"a}ube.
\newblock In {\em Diploma Dissert}. Technikum Winterthur, 1977.

\bibitem{bartknecht1989dust}
W.~Bartknecht and G.~Zwahlen.
\newblock {\em Dust explosions: course, prevention, protection}.
\newblock Springer, 1989.

\bibitem{pu1991turbulence}
Y.~Pu, J.~Jarosinski, V.~Johnson, and C.~Kauffman.
\newblock Turbulence effects on dust explosions in the 20-liter spherical
  vessel.
\newblock In {\em Symposium (International) on Combustion}, volume 23-1, pages
  843--849. Elsevier, 1991.

\bibitem{dahoe2001transient}
A.~Dahoe, R.~Cant, M.~Pegg, and B.~Scarlett.
\newblock On the transient flow in the 20-liter explosion sphere.
\newblock {\em Journal of Loss Prevention in the Process Industries},
  14(6):475--487, 2001.

\bibitem{van1992interpretation}
P.~Van~der Wel, J.~Van~Veen, S.~Lemkowitz, B.~Scarlett, and C.~Van~Wingerden.
\newblock An interpretation of dust explosion phenomena on the basis of time
  scales.
\newblock {\em Powder technology}, 71(2):207--215, 1992.

\bibitem{PROUST2007599}
C.~Proust, A.~Accorsi, and L.~Dupont.
\newblock Measuring the violence of dust explosions with the “20l sphere”
  and with the standard “{ISO} 1m3 vessel”: Systematic comparison and
  analysis of the discrepancies.
\newblock {\em Journal of Loss Prevention in the Process Industries},
  20(4):599--606, 2007.
\newblock Selected Papers Presented at the Sixth International Symposium on
  Hazards, Prevention and Mitigation of Industrial Explosions.

\bibitem{DISARLI2015204}
V.~{Di Sarli}, R.~Sanchirico, P.~Russo, and A.~{Di Benedetto}.
\newblock {CFD} modeling and simulation of turbulent fluid flow and dust
  dispersion in the {20-L} explosion vessel equipped with the perforated
  annular nozzle.
\newblock {\em Journal of Loss Prevention in the Process Industries},
  38:204--213, 2015.

\bibitem{vizcaya2018cfd}
D.~Vizcaya, A.~Pinilla, M.~Am{\'\i}n, N.~Ratkovich, F.~Munoz, C.~Murillo,
  N.~Bardin-Monnier, and O.~Dufaud.
\newblock {CFD} as an approach to understand flammable dust 20 {L} standard
  test: Effect of the ignition time on the fluid flow.
\newblock {\em AIChE Journal}, 64(1):42--54, 2018.

\bibitem{DIBENEDETTO2013cfd}
A.~Di~Benedetto, P.~Russo, R.~Sanchirico, and V.~Di~Sarli.
\newblock {CFD} simulations of turbulent fluid flow and dust dispersion in the
  20 liter explosion vessel.
\newblock {\em AIChE Journal}, 59(7):2485--2496, 2013.

\bibitem{di2014cfd}
V.~Di~Sarli, P.~Russo, R.~Sanchirico, and A.~Di~Benedetto.
\newblock {CFD} simulations of dust dispersion in the 20 {L} vessel: effect of
  nominal dust concentration.
\newblock {\em Journal of Loss Prevention in the Process Industries}, 27:8--12,
  2014.

\bibitem{KALEJAIYE201046}
O.~Kalejaiye, P.~R. Amyotte, M.~J. Pegg, and K.~L. Cashdollar.
\newblock Effectiveness of dust dispersion in the 20-{L Siwek} chamber.
\newblock {\em Journal of Loss Prevention in the Process Industries},
  23(1):46--59, 2010.

\bibitem{du2015visualization}
B.~Du, W.~Huang, L.~Liu, T.~Zhang, H.~Li, Y.~Ren, and H.~Wang.
\newblock Visualization and analysis of dispersion process of combustible dust
  in a transparent {Siwek} 20-{L} chamber.
\newblock {\em Journal of Loss Prevention in the Process Industries},
  33:213--221, 2015.

\bibitem{SANCHIRICO2015203}
R.~Sanchirico, V.~{Di Sarli}, P.~Russo, and A.~{Di Benedetto}.
\newblock Effect of the nozzle type on the integrity of dust particles in
  standard explosion tests.
\newblock {\em Powder Technology}, 279:203--208, 2015.

\bibitem{murillo2016cfd}
C.~Murillo, N.~Bardin-monnier, C.~Blanchard, D.~Funfschilling, F.~Munoz,
  N.~Rios, and D.~Vizcaya.
\newblock {CFD} to improve the repeatability and accuracy of dust explosion
  tests in the 20-liters sphere.
\newblock {\em Chemical Engineering Transactions}, 48:115--120, 2016.

\bibitem{SKJOLD2007291}
T.~Skjold.
\newblock Review of the {DESC} project.
\newblock {\em Journal of Loss Prevention in the Process Industries},
  20(4):291--302, 2007.
\newblock Selected Papers Presented at the Sixth International Symposium on
  Hazards, Prevention and Mitigation of Industrial Explosions.

\bibitem{FLACS}
Gexcon.
\newblock {\em FLACS Sofware}, 2021.
\newblock Bergen, Norway.

\bibitem{LI2020118028}
H.~Li, J.~Deng, C.-M. Shu, C.-H. Kuo, Y.~Yu, and X.~Hu.
\newblock Flame behaviours and deflagration severities of aluminium
  powder–air mixture in a 20-{L} sphere: Computational fluid dynamics
  modelling and experimental validation.
\newblock {\em Fuel}, 276:118028, 2020.

\bibitem{RAY2020321}
S.~K. Ray, N.~K. Mohalik, A.~M. Khan, D.~Mishra, N.~K. Varma, J.~K. Pandey, and
  P.~K. Singh.
\newblock {CFD} modeling to study the effect of particle size on dispersion in
  20l explosion chamber: An overview.
\newblock {\em International Journal of Mining Science and Technology},
  30(3):321--327, 2020.

\bibitem{WANG2019509}
S.~Wang, Z.~Shi, X.~Peng, Y.~Zhang, W.~Cao, W.~Chen, and J.~Li.
\newblock Effect of the ignition delay time on explosion severity parameters of
  coal dust/air mixtures.
\newblock {\em Powder Technology}, 342:509--516, 2019.

\bibitem{MURILLO2013103}
C.~Murillo, O.~Dufaud, N.~Bardin-Monnier, O.~López, F.~Munoz, and L.~Perrin.
\newblock Dust explosions: {CFD} modeling as a tool to characterize the
  relevant parameters of the dust dispersion.
\newblock {\em Chemical Engineering Science}, 104:103--116, 2013.

\bibitem{WELLER1998}
H.~G. Weller, G.~Tabor, H.~Jasak, and C.~Fureby.
\newblock A tensorial approach to computational continuum mechanics using
  object-oriented techniques.
\newblock {\em Computers in Physics}, 12(6):620--631, 1998.

\bibitem{LYSENKO2013408}
D.~A. Lysenko, I.~S. Ertesvåg, and K.~E. Rian.
\newblock Modeling of turbulent separated flows using {OpenFOAM}.
\newblock {\em Computers \& Fluids}, 80:408--422, 2013.
\newblock Selected contributions of the 23rd International Conference on
  Parallel Fluid Dynamics ParCFD2011.

\bibitem{TAVAKKOL2021110582}
S.~Tavakkol, T.~Zirwes, J.~A. Denev, F.~Jamshidi, N.~Weber, H.~Bockhorn, and
  D.~Trimis.
\newblock An {Eulerian-Lagrangian} method for wet biomass carbonization in
  rotary kiln reactors.
\newblock {\em Renewable and Sustainable Energy Reviews}, 139:110582, 2021.

\bibitem{launder1983numerical}
B.~E. Launder and D.~B. Spalding.
\newblock The numerical computation of turbulent flows.
\newblock In {\em Numerical prediction of flow, heat transfer, turbulence and
  combustion}, pages 96--116. Elsevier, 1983.

\bibitem{elghobashi1994predicting}
S.~Elghobashi.
\newblock On predicting particle-laden turbulent flows.
\newblock {\em Applied scientific research}, 52(4):309--329, 1994.

\bibitem{putnam1961integratable}
A.~Putnam.
\newblock Integratable form of droplet drag coefficient.
\newblock {\em Ars Journal}, 31(10):1467--1468, 1961.

\bibitem{crowe2011multiphase}
C.~T. Crowe, J.~D. Schwarzkopf, M.~Sommerfeld, and Y.~Tsuji.
\newblock {\em Multiphase flows with droplets and particles}.
\newblock CRC press, 2011.

\bibitem{haervig2017adhesive}
J.~H{\ae}rvig.
\newblock {\em On the Adhesive Behaviour of Micron-sized Particles in Turbulent
  Flow: A Numerical Study Coupling the Discrete Element Method and Large Eddy
  Simulations}, 2017.

\bibitem{PICO2020638}
P.~Pico, N.~Ratkovich, F.~Muñoz, and O.~Dufaud.
\newblock {CFD-DPM} and experimental study of the dynamics of wheat starch
  powder/pyrolysis gases hybrid mixtures in the 20-{L} sphere.
\newblock {\em Powder Technology}, 372:638--658, 2020.

\bibitem{gosman1983aspects}
A.~Gosman and E.~Loannides.
\newblock Aspects of computer simulation of liquid-fueled combustors.
\newblock {\em Journal of energy}, 7(6):482--490, 1983.

\bibitem{ranz1952evaporation}
W.~Ranz and W.~Marshall.
\newblock Evaporation from droplets.
\newblock {\em Chemical Engineering Progress}, 48(3):141--146, 1952.

\bibitem{baker2005mesh}
T.~J. Baker.
\newblock Mesh generation: Art or science?
\newblock {\em Progress in Aerospace Sciences}, 41(1):29--63, 2005.

\bibitem{celik2008procedure}
I.~B. Celik, U.~Ghia, P.~J. Roache, and C.~J. Freitas.
\newblock Procedure for estimation and reporting of uncertainty due to
  discretization in {CFD} applications.
\newblock {\em Journal of fluids Engineering-Transactions of the ASME}, 130(7),
  2008.

\bibitem{MURILLO201854}
C.~Murillo, M.~Amín, N.~Bardin-Monnier, F.~Muñoz, A.~Pinilla, N.~Ratkovich,
  D.~Torrado, D.~Vizcaya, and O.~Dufaud.
\newblock Proposal of a new injection nozzle to improve the experimental
  reproducibility of dust explosion tests.
\newblock {\em Powder Technology}, 328:54--74, 2018.

\bibitem{portarapillo2020cfd}
M.~Portarapillo, V.~Di~Sarli, R.~Sanchirico, and A.~Di~Benedetto.
\newblock {CFD} simulation of the dispersion of binary dust mixtures in the
  20{L} vessel.
\newblock {\em Journal of Loss Prevention in the Process Industries},
  67:104231, 2020.

\bibitem{balachandar2010turbulent}
S.~Balachandar and J.~K. Eaton.
\newblock Turbulent dispersed multiphase flow.
\newblock {\em Annual review of fluid mechanics}, 42:111--133, 2010.

\bibitem{ferrante2003physical}
A.~Ferrante and S.~Elghobashi.
\newblock On the physical mechanisms of two-way coupling in particle-laden
  isotropic turbulence.
\newblock {\em Physics of fluids}, 15(2):315--329, 2003.

\bibitem{kussin2002experimental}
J.~Kussin and M.~Sommerfeld.
\newblock Experimental studies on particle behaviour and turbulence
  modification in horizontal channel flow with different wall roughness.
\newblock {\em Experiments in Fluids}, 33(1):143--159, 2002.

\bibitem{yao2020analysis}
N.~Yao, L.~Wang, C.~Bai, N.~Liu, and B.~Zhang.
\newblock Analysis of dispersion behavior of aluminum powder in a 20 {L}
  chamber with two symmetric nozzles.
\newblock {\em Process Safety Progress}, 39(1):e12097, 2020.

\bibitem{DiSarli2013727}
V.~Di~Sarli, P.~Russo, R.~Sanchirico, and A.~Di~Benedetto.
\newblock {CFD} simulations of the effect of dust diameter on the dispersion in
  the 20 l bomb.
\newblock {\em Chemical Engineering Transactions}, 31:727--732, 2013.

\end{thebibliography}




\clearpage

\begin{table}
\caption{Mesh quality metrics reported by OpenFOAM's \texttt{checkMesh} utility}
\label{TAB:1}
\begin{tabular*}{\tblwidth}{@{} LLLL@{} }
\toprule
Parameter              & Value                                                                                                \\ \midrule
Max. Aspect Ratio           & 25.32                                                                                                 \\
Avg. Non-Orthogonality      & 13.72 (max 75.36)                                                                                      \\
Max. Skewness           & 1.68                                                                                                 \\
Avg. Cell determinant       & 3.81                                                                                                 \\
Avg. Face interpolation wt. & 0.47                                                                                                 \\
Number of cells        & \begin{tabular}[c]{@{}l@{}}Sphere vessel: 1,261,004 \\ Pipe: 313,500 \\ Canister: 39,710 \end{tabular} 
\\ \bottomrule
\end{tabular*}
\end{table}

\begin{table}
\caption{Boundary conditions case set-up}
\label{TAB:2}
\begin{tabular*}{\tblwidth}{@{} LLLL@{} }
\toprule
Variable   & Boundary Condition  & Initial Value \\
\midrule
U          & noSlip              & 0 m/s         \\
T          & fixedValue          & 293 K         \\
k          & kqRWallFunction     & 1  $\text{m}^{2}\text{s}^{-2}$ \\
$\varepsilon$   & epsilonWallFunction & 117 $\text{m}^{2}\text{s}^{-3}$   \\
$O_{2}/N_{2}$      & zeroGradient        & 0.23/0.77 (\%w)      \\
$p_{\text{sphere}}$ & zeroGradient        & 0.4 bar       \\ 
$p_{\text{canister}}$ & zeroGradient        & 21 bar       \\ 
\bottomrule
\end{tabular*}
\end{table}

\begin{table}
\caption{Calculations of numerical error introduced by spatial discretization using the GCI method.}
\label{Table:GCI_results}

\begin{tabular}{@{}llllll@{}}
\toprule
 & \multicolumn{1}{c}{\begin{tabular}[c]{@{}c@{}}$\phi$ pressure (bar) \\ in the sphere at $60\text{ms}$ \end{tabular}} & \multicolumn{1}{c}{\begin{tabular}[c]{@{}c@{}}$\phi$ TKE $\left(\text{m}^2\text{s}^{-2}\right)$ \\ in the sphere at $60\text{ms}$ \end{tabular}} &  & \multicolumn{1}{c}{\begin{tabular}[c]{@{}c@{}}$\phi$ pressure (bar) \\ in the sphere at $60\text{ms}$\end{tabular}} & \multicolumn{1}{c}{\begin{tabular}[c]{@{}c@{}}$\phi$ TKE $\left(\text{m}^2\text{s}^{-2}\right)$ \\ in the sphere  at $60\text{ms}$\end{tabular}} \\ \midrule
$N_{2}, N_{3}, N_{4}$ & \multicolumn{2}{c}{3.25M, 1.62M, 0.83M} & $N_{1}, N_{2}, N_{3}$ & \multicolumn{2}{c}{7.5M, 3.25M, 1.62M} \\
$r_{32}$ & \multicolumn{2}{c}{1.26} & $r_{21}$ & \multicolumn{2}{c}{1.32} \\
$r_{43}$ & \multicolumn{2}{c}{1.25} & $r_{32}$ & \multicolumn{2}{c}{1.26} \\
$\phi_{2}$ & \multicolumn{1}{c}{0.9807} & \multicolumn{1}{c}{86.121} & $\phi_{1}$ & \multicolumn{1}{c}{0.9810} & \multicolumn{1}{c}{82.612}  \\
$\phi_{3}$ & \multicolumn{1}{c}{0.9823} & \multicolumn{1}{c}{79.639} & $\phi_{2}$ & \multicolumn{1}{c}{0.9807} & \multicolumn{1}{c}{86.121}  \\
$\phi_{4}$ & \multicolumn{1}{c}{0.9829} & \multicolumn{1}{c}{64.527} & $\phi_{3}$ & \multicolumn{1}{c}{0.9823} & \multicolumn{1}{c}{79.639}  \\
$p$ & \multicolumn{1}{c}{4.52} & \multicolumn{1}{c}{3.75} & $p$ & \multicolumn{1}{c}{6.9}  & \multicolumn{1}{c}{2.48}  \\
$\phi_{\text{ext}}^{32}$ & \multicolumn{1}{c}{0.9798}  & \multicolumn{1}{c}{90.83} & $\phi_{\text{ext}}^{21}$ & \multicolumn{1}{c}{0.9810} & \multicolumn{1}{c}{79.08}  \\
$e_{a}^{32}$, $e_{\text{ext}}^{32}$ & \multicolumn{1}{c}{0.17\%, 0.09\%}  & \multicolumn{1}{c}{7.5\%, 5.1\%} & $e_{a}^{21}$, $e_{\text{ext}}^{21}$ & \multicolumn{1}{c}{0.03\%, 0.005\%} & \multicolumn{1}{c}{4.2\%,4.4\%}  \\
$GCI_{\text{fine}}^{32}$ & \multicolumn{1}{c}{0.11\%}  & \multicolumn{1}{c}{6.83\%} & $GCI_{\text{ultrafine}}^{21}$ & \multicolumn{1}{c}{0.007\%}  & \multicolumn{1}{c}{5.33\%}  \\ \bottomrule
\end{tabular}

\end{table}

\clearpage

\clearpage

\clearpage

\clearpage

\clearpage

\clearpage

\clearpage

\end{document}